\documentclass{aa}  

\usepackage{graphicx}
\usepackage[labelfont=bf]{subcaption}
\usepackage{txfonts}
\usepackage{mathtools}

\usepackage{rotating}

\usepackage{hyperref} 
\usepackage{color}

\newcommand{\G}{\mathcal{G}}
\newcommand{\Msun}{M_\odot}
\newcommand{\Rsun}{R_\odot}
\newcommand{\Ms}{M_{\star}}
\newcommand{\Rs}{R_{\star}}
\newcommand{\Ts}{T_\star}
\newcommand{\Mp}{M_{\rm p}}

\newcommand{\dd}{\mathrm{d}}
\newcommand{\Os}{\Omega_{\star}}
\newcommand{\sss}{\sigma_{\star}}
\newcommand{\oss}{\overline{\sss}}

\newcommand{\Qs}{\overline{Q_s'}}

\begin{document}

   \title{Tidal dissipation in rotating low-mass stars \\ 
   and implications for the orbital evolution of close-in massive planets}
   \subtitle{II. Effect of stellar metallicity}

\titlerunning{Effect of stellar metallicity}

\author{E. Bolmont\inst{1,2} \and F. Gallet\inst{3}  \and S. Mathis\inst{1}  \and C. Charbonnel\inst{3,4} \and L. Amard\inst{3,5} \and Y. Alibert\inst{6}}

\offprints{E. Bolmont,\\ email: emeline.bolmont@cea.fr}

\institute{ 
$^1$ Laboratoire AIM Paris-Saclay, CEA/DRF - CNRS - Univ. Paris Diderot - IRFU/SAp, Centre de Saclay, F-91191 Gif-sur-Yvette Cedex, France \\
$^2$ NaXys, Department of Mathematics, University of Namur, 8 Rempart de la Vierge, 5000 Namur, Belgium \\
$^3$ Department of Astronomy, University of Geneva, Chemin des Maillettes 51, 1290 Versoix, Switzerland \\
$^4$ IRAP, UMR 5277, CNRS and Universit\'e de Toulouse, 14, av. E.Belin, F-31400 Toulouse, France \\
$^5$ LUPM UMR 5299 CNRS/UM, Universit\'e de Montpellier, CC 72, F-34095 Montpellier Cedex 05, France \\
$^6$ Physikalisches Institut \& Center for Space and Habitability, Universitaet Bern, 3012 Bern, Switzerland}

  \date{Submitted to A\&A}%, please \NB{DO NOT CIRCULATE}}
%   \date{Received --; accepted --}

% \abstract{}{}{}{}{} 
% 5 {} token are mandatory
 
  \abstract
   {Observations of hot Jupiter type exoplanets suggest that their orbital period distribution depends on the metallicity of their host star.
We investigate here whether the impact of the stellar metallicity on the evolution of the tidal dissipation inside the convective envelope of rotating stars and its resulting effect on the planetary migration might be a possible explanation for this observed statistical trend.

We use a frequency-averaged tidal dissipation formalism coupled to an orbital evolution code and to rotating stellar evolution models in order to estimate the effect of a change of stellar metallicity on the evolution of close-in planets. 
We consider here two different stellar masses: 0.4 and 1.0 $M_{\odot}$ evolving from the early pre-main sequence phase up to the red giant branch.

We show that the metallicity of a star has a strong effect on the stellar parameters which in turn strongly influence the tidal dissipation in the convective region. While on the pre-main sequence the dissipation of a metal poor Sun-like star is higher than the dissipation of a metal rich Sun-like star, on the main sequence it is the opposite. However, for the $0.4~\Msun$ star, the dependence of the dissipation with metallicity is much less visible.

Using an orbital evolution model, we show that changing the metallicity leads to different orbital evolutions (e.g., planets migrate farther out from an initially fast rotating metal rich star). By using this model, we qualitatively reproduced the observational trends of the population of hot Jupiters with the metallicity of their host stars. However, more steps are needed to improve our model to try to quantitatively fit our results to the observations. Namely, we need to improve the treatment of the rotation evolution in the orbital evolution model and ultimately we need to consistently couple of the orbital model to the stellar evolution model.}

   \keywords{Planet-star interactions -- Stars: evolution -- Stars: rotation -- Stars: abundances -- Stars: solar-type}

   \maketitle
%
%________________________________________________________________

\section{Introduction}

%and the future ones as TESS \citep{TESS}, Plato \citep{Plato}, and SPIRou \citep{Spirou}. 

%Thanks to the Kepler and K2 missions (\textbf{ADD others)+ref}

Thanks to space missions such as CoRoT \citep{Corot}, Kepler \citep{Kepler}, and K2 \citep{K2}, the amount of detected and confirmed exoplanets reaches an exquisite high number of objects belonging to an expended range of star-planet system's configurations. 
The high number of detected exoplanets allows us to do some statistics over the characteristics of the host star. 
For instance, observational studies such as \citet{Gonzalez1997}, \citet{Santos2003} or \citet{Adibekyan13} have shown that there are some features of the exoplanet population which depend on the metallicity of the host star.
Observational surveys showed that there seems to be more planets around metal-rich stars than around metal poor stars \citep[]{Gonzalez1997,Santos2003,Neves2013}.
Indeed, metal rich stars host metal rich disks which have more material for planet formation.
\citet{BN13} found that not only the number of planets depends on the stellar metallicity but also the mass and radius of the planets, namely that there are fewer small size/low-mass planets around metal-poor stars than around metal-rich stars.
Furthermore, studies such as the works of \citet{Adibekyan13} have shown that planets between 10 $M_{\oplus}$ and 4 $M_{\rm jup}$ orbiting around metal-poor stars have on average a wider semi-major axis (longer orbital period) than those orbiting metal-rich stars.
More specifically, the last two aspects appear to be true for the hot Jupiter population.

The origin of this metallicity dependence can be due to several aspects: the formation processes, their efficiency and the disk driven migration might be dependent on the metallicity of the host star and its disk \citep[especially type I migration, which strongly depends on the thermodynamics and thus on the opacity of the disk, e.g.][]{Mordasini2009}, the observations might be biased in favor of planets around metal-rich stars \citep{Santos2005}, the tidal orbital evolution of the close-in massive planets might be dependent on metallicity via the effect of the metallicity on the structural and rotational properties and thus the dissipation of the star.
This article aims at investigating the implications of the latter proposition.

Stars are born inside a molecular cloud which  metallicity depends on the region of the Galaxy where it is located. 
A change of the stellar metallicity, a quantity that estimates the fraction of mass of atoms heavier than helium, has a strong effect on the overall stellar properties and modifies the size, the internal structure, and the lifetime of a star of a given initial mass. 
For instance the size of the convective envelope of a low-mass star at a given evolution phase decreases %for a decreasing 
with metallicity due to the facilitated energy redistribution which leads to a decrease of the opacity and an increase of the stellar effective temperature (see Fig. \ref{stellarstructure}); as a consequence, the dissipation of tidal waves inside the envelope will be reduced as well (e.g. \citealt{Mathis15} and \citealt{Gallet16}, hereafter Paper I). 
Thus, two star-planet systems having all characteristics equal except for the metallicity of the host star would not necessary experience the same tidal evolution. 

%While the main physical processes of the interaction between star and planet are now fairly well understood, detailed analysis is now required so as to account for the new and more detailed observations. 
%More specifically, tidal interaction between rotating stars and their orbiting planet(s) have to be more intensively investigated and sophisticated. 
Thanks to recent progress made in the modeling of the dynamical tide induced dissipation occurring in the convective envelope of rotating stars \citep[the frequency-averaged tidal dissipation, which depends on the stellar internal structure properties and surface rotation rate, see][]{Ogilvie13,Mathis15}, % have allowed us to 
we could revisit the tidal evolution of close-in planets \citep[][hereafter Paper I]{Bolmont16,Gallet16}.
Contrary to the equilibrium tide which is a large scale flow induced by the hydrostatic readjustment of the stellar structure caused by a perturbing potential \citep[e.g.][]{Zahn66}, the dynamical tide in the convective envelope consists in inertial waves excited by the perturber and driven by the Coriolis acceleration \citep[e.g.][]{Ogilvie07}.
Both phenomena lead to dissipation inside the star which impact the orbit of the perturber and the rotation of the star.
%Tidal dissipation inside the convective envelope of rotating stars have recently benefited from a \FG{averaged} analysis of \citet{Ogilvie13} and \citet{Mathis15} in which they expressed this dissipation as a function of the stellar internal structure properties and surface rotation rate. 
In %\citet{Gallet16}, hereafter 
Paper I, we primarily investigated the effect of the stellar mass (and thus internal structure) on the frequency-averaged tidal dissipation and modified equivalent tidal quality factor from the pre-main sequence (hereafter PMS) up to the red-giant branch (hereafter RGB). 
In this article we investigate the effect of a change of the stellar metallicity both on the frequency-averaged tidal dissipation using the stellar evolution code STAREVOL \citep[e.g.][, which includes rotation]{Amard15}, and on the resulting tidal orbital evolution of close-in hot Jupiters using the orbital evolution model of \citet{Bolmont16}.

%Moreover, the mass and radius of planets depend on the stellar metallicity as pointed out by \citet{BN13} that found a lack of small size/low-mass planets around metal-poor stars. Interestingly, and by using observations of star-planet(s) systems, \citet{Adibekyan13} found that planets between 10 $M_{\oplus}$ and 4 $M_{jup}$ orbiting around metal-poor stars have on average a wider semi-major axis (longer orbital period) than those orbiting metal-rich stars.

%Metallicity affect the internal structure of the star (when metallicity decreases size of the radiative core increases, which reduced the dissipation of tidal waves inside the convective envelope of the stars).

%Moreover the mass and radius of planets depend on the stellar metallicity  \citep[][that found a lack of small size/low-mass planets around metal-poor stars]{BN13}.

%Due to an high abundance of heavy element, metal rich stars are more yield to host massive planet (\textbf{CITATION}) compare to metal poor star. The stellar metallicity is then an indirect but strong limiting factor to the formation processes of Hot Jupiter type planets.
%\textbf{Cite \citet{Adibekyan13}}.

%The aim of the work is to investigate the impact of the stellar metallicity on the evolution of the \FG{frequency-averaged} tidal dissipation and \FG{modified equivalent} quality factor of rotating star using the stellar structure evolution code STAREVOL. 

The article is organized as follows: in Sect.~\ref{model} we describe the stellar and tidal models. In Sect.~\ref{metaleffect} we discuss the effect of the metallicity of the star on its structural parameters as well as on its tidal dissipation. 
In Sect.~\ref{orbits} we discuss the implications of the dependence of the dissipation with metallicity for the tidal orbital evolution of close-in massive planets and also the implications of our study on the understanding of the statistics of exoplanets. 
Finally, we conclude in Sect.~\ref{conc}.

\section{Tidal star-planet interaction model}
\label{model}

The description of the model is already detailed in Paper I. 
We recall here its main features.

%\subsection{Tidal evolution model}

The tidal interactions between a star and a short-period planet depend on the equilibrium and the dynamical tides in the host star \citep[e.g.][]{Zahn66,Zahn75,Zahn77,MathisRemus13,Ogilvie14}. 
The equilibrium tide and the dynamical tide are two different types of response of a fluid body to a perturbing potential.
On the one hand, the equilibrium tide is due to the hydrostatic adjustment of the stellar structure and the induced large-scale flow caused by the presence of the planet \citep{Zahn66,Remus2012,Ogilvie13}. 
On the other hand, the dynamical tide is caused by 1) the propagation inside the stellar convective envelope of inertial waves driven by the Coriolis acceleration and excited when the tidal frequency $\omega = 2\vert n - \Omega_\star\vert$\footnote{For a circular, non inclined orbits.} ranges between $[-2\Omega_{\star}, 2\Omega_{\star}]$, where $\Omega_{\star}$ is the stellar spin and $n$ the orbital frequency of the planet \citep{Ogilvie07}; and 2) the internal gravity waves in the radiative core, which become gravito-inertial waves when affected by the Coriolis acceleration \citep[][]{Zahn75,Zahn77,Berthomieu1978,Goldreich892,Goldreich89,Terquem98,BarkerOgilvie2010}. 
The dynamical tide strongly depends on the excitation frequency, the age and the rotation rate of the star \citep[see the discussion in][]{Ogilvie04,Ogilvie07,AD15,Witte02}.
Hence, its behavior is unfortunately quite complex and thus too time consuming to compute for a complete evolution of the system and do a wide exploration of systems configurations \citep[see][]{Bolmont16}. 

In this work, we neglect the tidal dissipation in the radiative region of the star and only consider 
the tidal dissipation in the convective envelope.
We also consider circular and non-inclined orbits.\\
As in \citet{Bolmont16} and \cite{Gallet16}, we use the two-layer simplified model introduced by \citet{Ogilvie13} and \citet{Mathis15} to evaluate the frequency-averaged tidal dissipation $<\mathcal{D}>_{\omega}$ in the stellar convective envelopes:
\begin{flalign}
\label{dissipequa}
<\mathcal{D}>_{\omega} =& \int^{+\infty}_{-\infty} {\rm Im}\left[k_2^2(\omega)\right] \frac{d\omega}{\omega} = \epsilon^2 \times g(\alpha, \beta),
\end{flalign}
where ${\rm Im}\left[k_2^2(\omega)\right]$ is the imaginary part of the Love number corresponding to the quadrupolar mode ($k_m^l$, when l = 2, m = 2, see \citealt{Ogilvie13}). 
$<\mathcal{D}>_{\omega}$ can be decomposed in a product of $\epsilon^2$ where $\epsilon = \left(\Omega_\star/\sqrt{\G M_\star /R_\star^3}\right)$ is the stellar rotation normalized by the Keplerian critical velocity and a function $g(\alpha, \beta)$ of the stellar structure.
The expression of $g(\alpha, \beta)$ is given by Eqs.~1 and 2 in Paper I.
Here $\alpha = R_c/\Rs$ and $\beta = M_c/\Ms$ are respectively the stellar radius aspect ratio and mass aspect ratio.
$R_c$ is the radius of the radiative core and $M_c$ its mass.

\medskip

To compute the dissipation, we therefore need to know the evolution of the structural parameters of the stars and of their rotation rate. 
Following the approach presented in paper I, we use the code STAREVOL \citep[see e.g.][]{Amard15} for 0.4 and 1.0~M$_{\odot}$ stars at three metallicities;  solar metallicity 
(Z = 0.0134, \citealt{AsplundGrevesse2009}), sub-solar metallicity 
(Z = 0.004), and super-solar metallicity 
(Z = 0.0255), corresponding respectively to [Fe/H]=-0.5, 0.0, and +0.3.
Rotating stellar models allow us to obtain those quantities for the different stellar evolution stages, from the PMS to the RGB phases.

\medskip

Once the frequency averaged dissipation $<\mathcal{D}>_{\omega}$ is computed, we use the formalism of \citet{Bolmont16} to compute the orbital evolution of a given planet.
The secular evolution of the semi-major axis $a$ of a planet on a circular orbit is given by \citep{Hansen10,Leconte2010,Bolmont11,Bolmont12}: 
\begin{equation}\label{Hansena}
\frac{1}{a}\frac{\dd a}{\dd t} = - \frac{1}{\Ts}\Big[1-\frac{\Os}{n}\Big],
\end{equation}
where $\Ts$ is an evolution dissipation timescale given by:
\begin{equation}
\label{Ts}
\Ts = \frac{1}{9}\frac{\Ms}{\Mp(\Mp+\Ms)}\frac{a^8}{\Rs^{10}}\frac{1}{\sigma_{\star}},
\end{equation}
which depends on the semi-major axis $a$ and mass $\Mp$ of the planet, on the mass $\Ms$ and radius $\Rs$ of the star and a factor $\sigma_{\star}$, which we call the stellar dissipation factor.

If the tidal frequency $\omega$ is out of the range $[-2\Os, 2\Os]$, the equilibrium tide drives the evolution and the dissipation factor is taken to be the equilibrium tide dissipation factor as determined by \citet{Hansen12}.
If $\omega$ is in the range $[-2\Os, 2\Os]$, the planet excites the tidal inertial waves in the stellar convective layer and the dissipation factor $\sigma_{\star}$ is given by:
\begin{equation}\label{ss}
\sss = \frac{1}{3}\frac{\G}{\Rs^5}\frac{1}{|n-\Os|}<\mathcal{D}>_{\omega},
\end{equation}
where $<\mathcal{D}>_{\omega}$ is given by Eq.~\ref{dissipequa}.
In the following, we use a normalized dissipation factor $\oss = \sss/\sigma_0$, where $\sigma_0 = \sqrt{\G/(\Msun\Rsun^7)}$.
%\EB{Add here definition of $\oss$!}
Given the dependency of the evolution timescale on semi-major axis, stellar radius and dissipation factor (see Eq.~\ref{Ts}), we see that the farther the planet, the smaller the radius of the star and the smaller the dissipation factor $\sigma_{\star}$, the higher the evolution timescale.

\medskip

As explained in Paper I, %\citet{Gallet16}, 
our stellar evolution and orbital evolution models are not strictly coupled. 
We use grids of the structural tidal quality factor $\Qs$ \citep[$\propto  \epsilon^{2}<\mathcal{D}>_{\omega}^{-1}$, see Eq. 6 of][]{Gallet16}, which are computed with a stellar evolution model assuming an initial rotation period of 1.4~day.
We then compute the rotation taking into account the tidal torque and the torque induced by the stellar winds \citep[see Eq. 14 of][which used a prescription for the stellar wind induced angular momentum loss rate from \citealt{Bouvier1997}]{Bolmont16}.
From this rotation, we can then calculate the frequency-averaged tidal dissipation $<\mathcal{D}>_{\omega}$ and use Equations \ref{Hansena} to \ref{ss} to compute the orbital evolution.

This means that the structural tidal quality factor we use for the orbital evolution model accounts for a part of the effects of rotation. 
More precisely, the direct effect of rotation on the structure of the star via the centrifugal acceleration is taken into account in STAREVOL and therefore in $\Qs$. 
Additionally, the indirect effect of rotation on the stellar structure due to the modification of its chemicals stratification because of the internal secular angular momentum transport and the induced mixing in the radiative core is also taken into account in STAREVOL and thus in $\Qs$. 
These two effects are however not treated consistently in the orbital evolution code where the star is simply considered as a uniformly rotating solid body.\\

Moreover, the evolution of the rotation differs slightly in the two models as the stellar wind prescription is different.
Indeed, the orbital evolution code uses the wind prescription from \citet{Bouvier1997} while the stellar evolution model uses the one of \citet{Matt15}.
Besides, in the stellar evolution models, the rotation is not solid as assumed in the orbital model: in particular, the radiative core rotates with a different frequency than the convective envelope \citep{Amard15}.
These differences in rotation calculation have two consequences. 
The first one is the effect of rotation on the structural tidal quality factor (via the structure and transport, see the previous paragraph). 
Moreover, tidal dissipation models taking into account this differential rotation should be built in a near future \citep[e.g.][]{Guenel2016a, Guenel2016b}. 
The second one is the effect of the wind-driven rotation evolution on the corotation radius and therefore on the orbital evolution of planets.

The first effect is negligible for rotations within the range we study here.
For instance, on the MS, the radius aspect ratio of an initially fast rotating 1~$\Msun$ star is less than 1\% bigger than the radius aspect ratio of a non-rotating star. 
This difference does not impact significantly the structural dissipation.
The second one can be important for the survival of inner planets, but this would not impact our overall results about the effect of metallicity.

The next steps for this work will be to implement the same stellar wind prescription in the orbital evolution model than in the stellar evolution code, consider the differential rotation between radiative core and envelope (e.g., using a two-layer model as in \citealt{GB13, GB15}) in the stellar and tidal models, and ultimately to consistently couple the orbital model with the rotating stellar evolution code.

\section{The effect of metallicity}
\label{metaleffect}

\begin{figure}
\begin{center}
\includegraphics[angle=-90,width=0.5\textwidth]{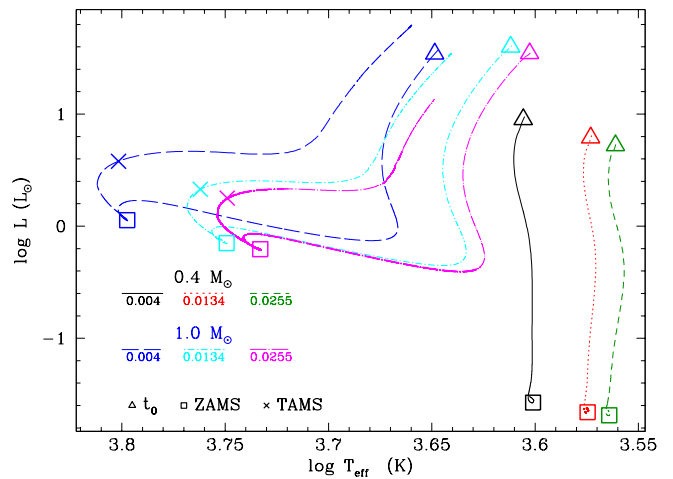}
\end{center}
\caption{Stellar evolution tracks in the Hertzsprung-Russell diagram for the rotating models of the 0.4 and 1.0 $M_{\odot}$ stars for the three metallicities considered. The symbol represent: the first step in each model (triangle), the zero-age main sequence (ZAMS, square), and the terminal-age main sequence (TAMS, cross).}
\label{hrd}
\end{figure}

When the metallicity of a star of a given mass is reduced, its opacity decreases (due to an easier redistribution of energy inside the star); this leads to a denser and hotter radiative core, and so to an increase of the luminosity and effective temperature of the star at a given evolution phase. 
One of the consequences of this easier redistribution of energy is that the star will burn its hydrogen reservoir more rapidly, thus reducing its overall lifetime. 
The metallicity of a star is directly linked to the metallicity of its parent molecular cloud. 
The latter provides the material (gas and dust) that will be used for the formation of the star and its surrounding disk. Given this initial amount of foundation bricks, the resulting star-disk systems diversity will range from a massive star surrounded by a thin disk to a low-mass star surrounded by a thick disk. 
As planets form inside the circumstellar disk, metallicity is also thought to impact their formation and migration processes \citep[e.g.][]{IdaLin2004,Mordasini2009,Alibert2013}.
Metallicity will then play an important role by indirectly setting the maximum mass of the planets, and possibly impacting also the location of their formation inside the circumstellar disk \citep[e.g.][]{Johnson2012,Mordasini12}.

\begin{figure*}[!ht]
\begin{center}
\includegraphics[angle=-90,width=0.4\textwidth]{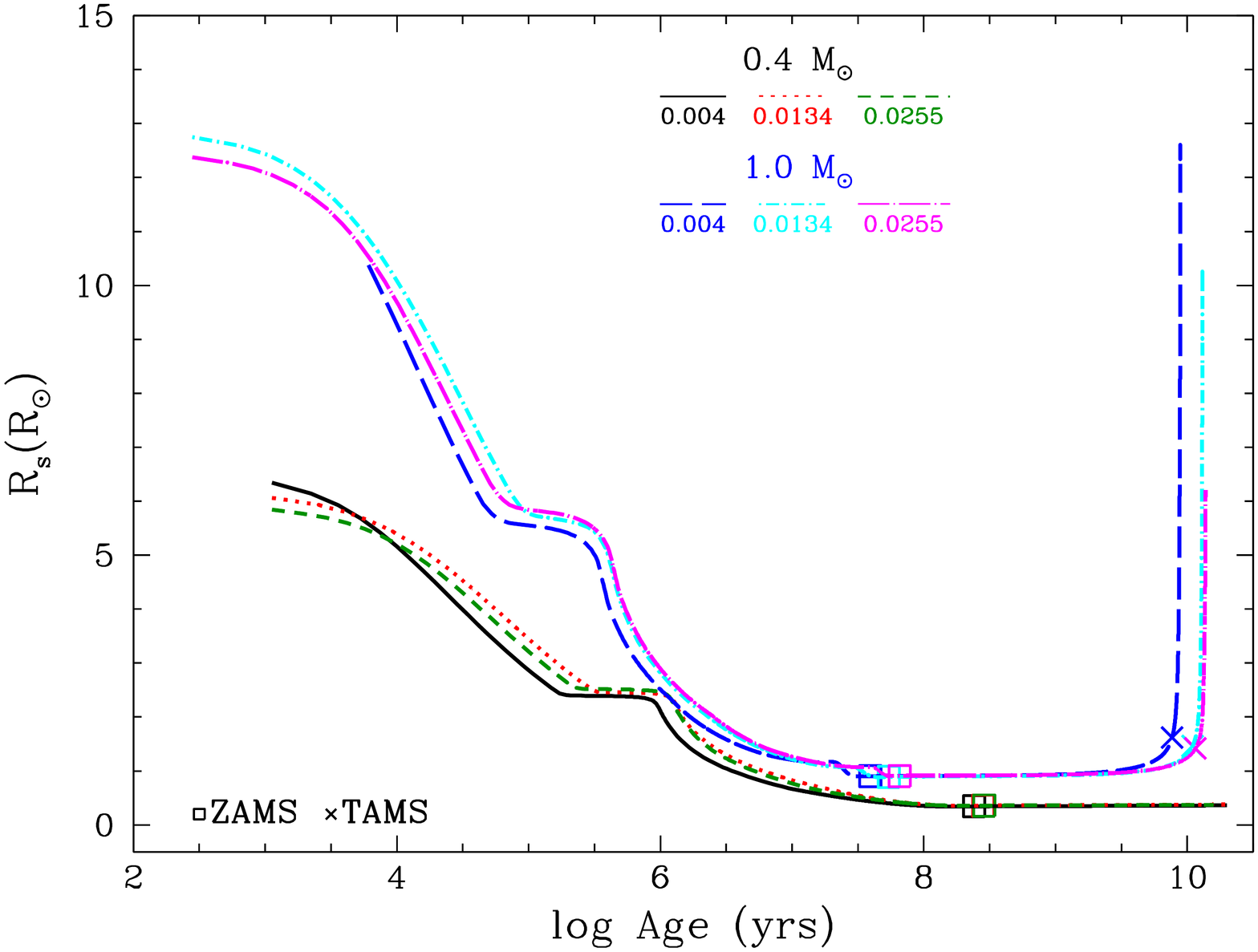}  \\
\includegraphics[angle=-90,width=0.4\textwidth]{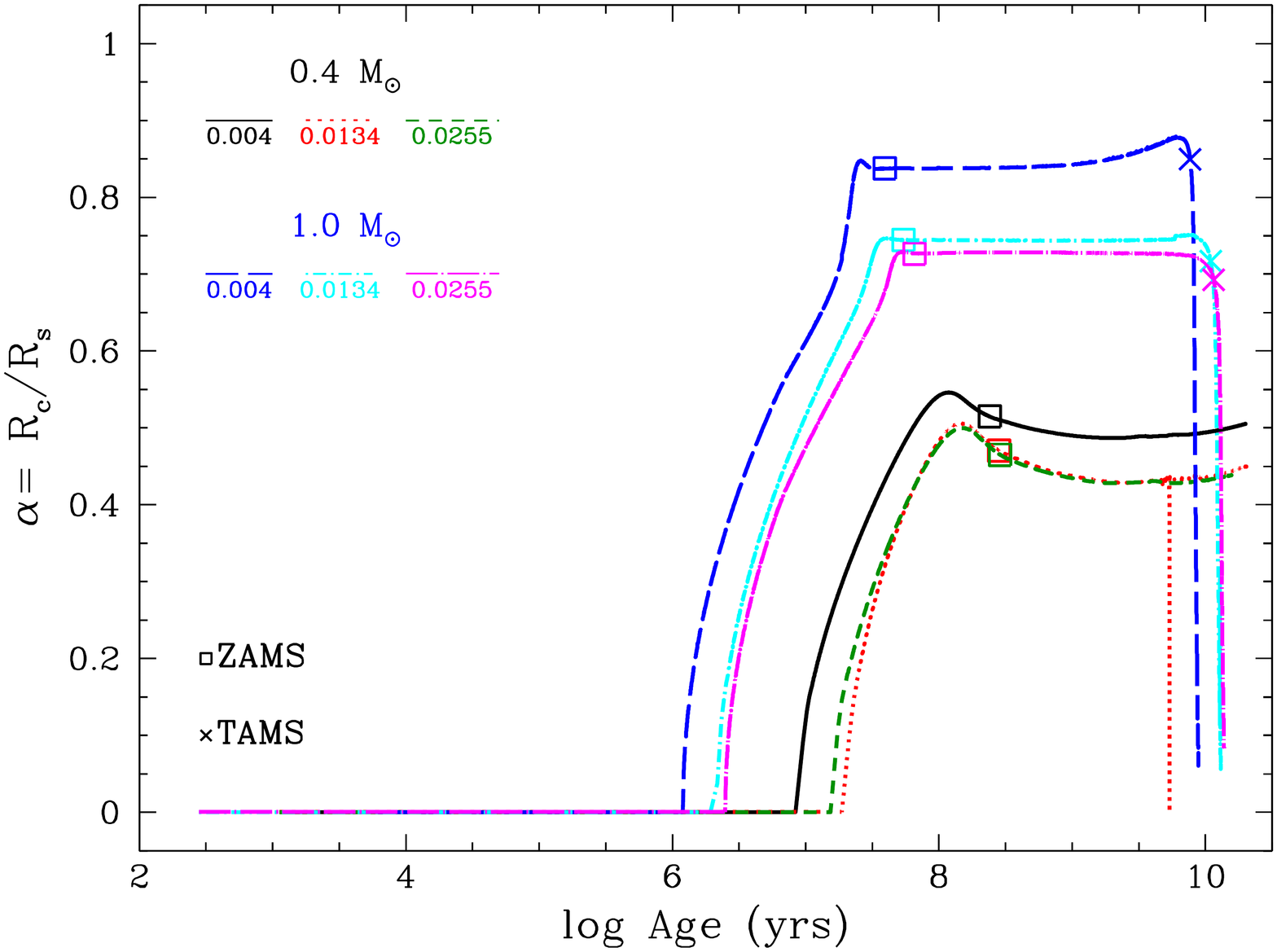} \hspace{0.1cm} \includegraphics[angle=-90,width=0.4\textwidth]{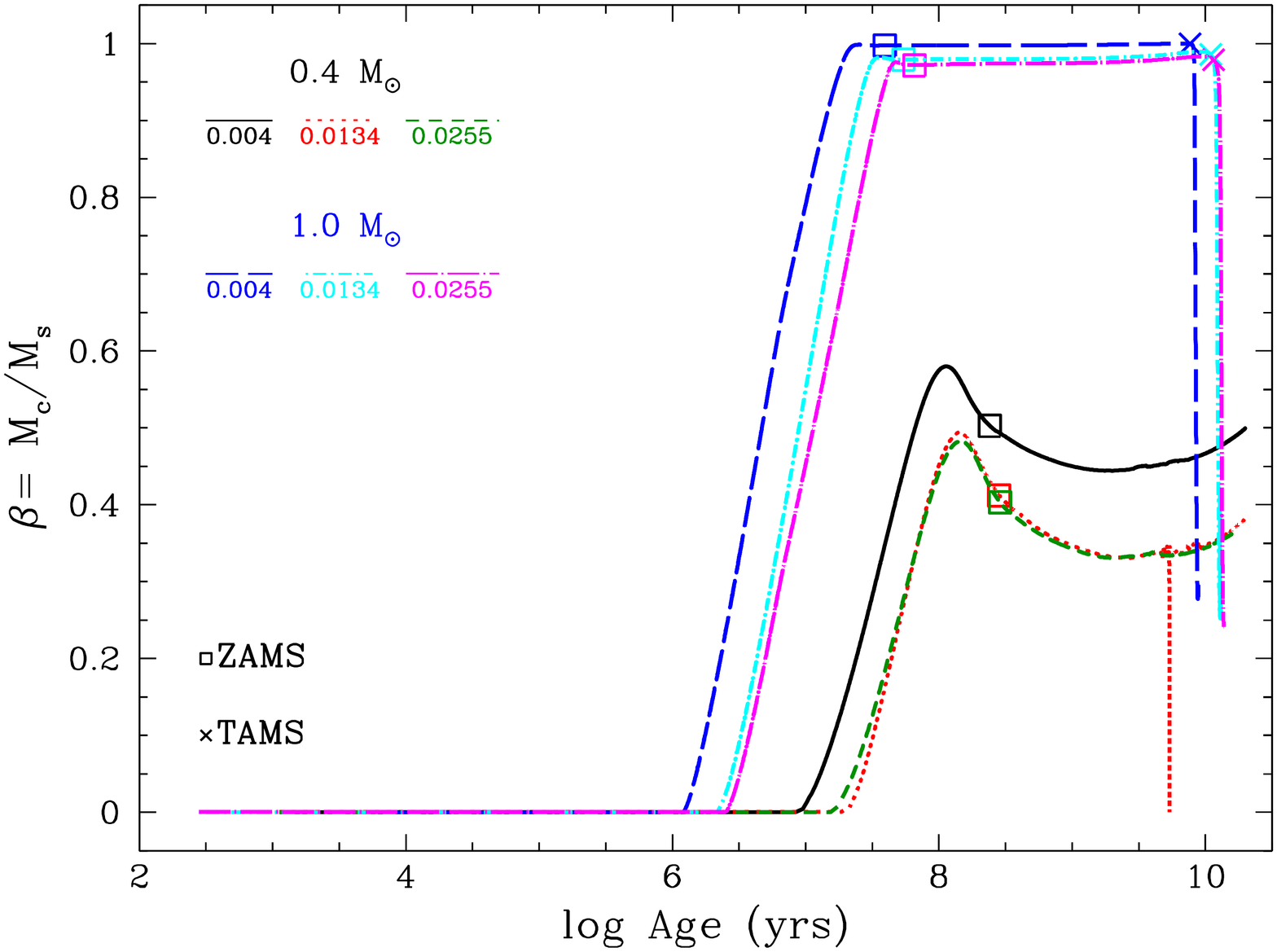}  
\end{center}
\caption{\textbf{Top:} Evolution of the stellar radius $R_{\star}$ of stars of 0.4 and 1.0 $M_{\odot}$ as a function of time and for the three metallicities considered. \textbf{Bottom:} Evolution of the radius aspect ratio $\alpha=R_{\rm{c}}/R_{\star}$ (left panel) and mass aspect ratio $\beta=M_{\rm{c}}/M_{\star}$ (right panel) of stars of 0.4 and 1.0 $M_{\odot}$ as a function of time and for the three metallicities considered. The symbol represent: the first step in each model (triangle), ZAMS (square), and TAMS (cross).}
\label{stellarstructure}
\end{figure*}

\subsection{Stellar evolution}

The effect of a change of metallicity on the stellar evolution tracks in the Hertzsprung-Russell Diagram (hereafter HRD) is illustrated in Fig.~\ref{hrd} for 0.4 and 1.0 $M_{\odot}$ stars with the three considered metallicities (Z=0.004, Z=0.0134=$Z_{\odot}$, and Z=0.0255).
As expected, the first striking effect of a decrease in metallicity is that at a given evolution phase the stars reach a higher effective temperature and luminosity due to a global decrease of their opacity.

Figure~\ref{stellarstructure} exhibits the evolution of the internal structure and stellar radius for the considered stars. 
It shows that metallicity is a key physical feature that influences significantly the structure of the star as well as important milestones in the evolution of the star such as the appearance of the radiative core on the PMS.
For instance, the appearance of the radiative core on the PMS occurs at an earlier age when the metallicity is lower. 

\subsubsection{Stellar radius}

The effect of metallicity on the stellar radius is shown in Fig.~\ref{stellarstructure} (top panel). 
Let us first consider the 1.0 $M_{\odot}$ models. 
During the early-PMS (i.e. between $10^4$ years and 0.5 Myr) the radius of the Z=0.0255 model is slightly smaller by about 6\% compared to the radius of the Z=0.0134 model.
Then from 0.5 Myr to up to the ZAMS the behavior reverses, the radius of the Z=0.0134 star becomes smaller by about 2\% compared to the radius of the Z=0.0255 star. 
During this whole period, the stellar radius of the Z=0.004 star is smaller by about 15\% compared to the radius of the Z=0.0255 star.
At their respective ZAMS age, the radius for the metal poor 1.0 $M_{\odot}$ model is slightly bigger than for the two other ones. 
All along the MS, this radius discrepancy between the metal poor star and the others increases to reach a maximum at the end of the MS.
For instance, at the age of the TAMS for the metal poor star model (at 7.64~Gyr), the stellar radius of the $Z=0.004$ star is about 40\% higher than that of the Z=0.0134 star and 50\% higher than that of the Z=0.0255 star.

On the other hand, the radius of the 0.4~$M_\odot$ star does not change as much since the evolution is much slower than for the 1.0~$M_\odot$ star. Globally, the more metal poor the star, the faster it will expand but also the smaller the radius at the ZAMS.

\begin{figure*}
\begin{center}
\includegraphics[angle=-90,width=0.5\textwidth]{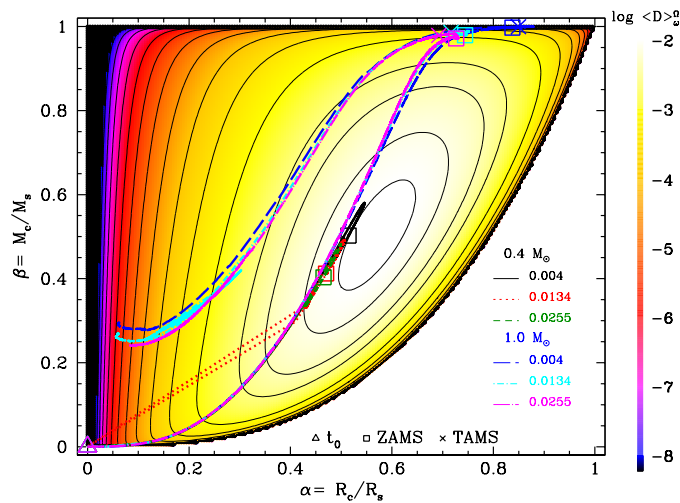}
\end{center}
\caption{Variation of $<\mathcal{D}>_{\omega}^{\Omega}=\epsilon^{-2}<\rm{Im}\left[k_2^2(\omega)\right]>_{\omega}$ as a function of aspect and mass ratio ($\alpha$ and $\beta$, respectively) in color scale. Evolutionary tracks of stars of 0.4 to 1.0 $M_{\odot}$ for the three metallicities considered in the ($\alpha , \beta$) plane. Levels are for $\log <\mathcal{D}>_{\omega}^{\Omega}=$ \{-2, -2.1, -2.3, -2.5, -3, -3.5, -4, -4.5, -5, -5.5, -6, -6.5, -7, -7.5, -8\}. The symbol represent: the first step in each model (triangle), ZAMS (square), and TAMS (cross).}
\label{chemin}
\end{figure*}

\subsubsection{Stellar structure}

The effect of metallicity on the radius and mass aspect ratios, $\alpha$ and $\beta$, is shown in Fig. \ref{stellarstructure} (bottom panel). For the 1 $M_{\odot}$ star at 1 Gyr, the $\alpha$ parameter for the metal-poor model is smaller by about 18\% than for the metal-rich one. 
This difference remains from the %first 
apparition of the radiative core around 1 Myr up to the end of the MS. $\beta$ follows the same evolution but with a difference of about 1\% between the two extreme metallicities. 
These behaviors are explained by a decrease of the averaged opacity of the star and an increase of its surface effective temperature as stellar metallicity decreases. Indeed, the easier redistribution of energy inside the star with decreasing metallicity allows the stellar interior to reach higher temperatures and to reach them faster than in metal rich stars. As a consequences, the radiative core of the 1 $M_{\odot}$ metal poor star starts to develop earlier along the evolution (around 1.2 Myr for the Z=0.004 case compared to 1.9 Myr and 2.5 Myr, for the Z=0.0134 and Z=0.0255 cases, respectively) and reaches higher radius ($\alpha$) and mass ($\beta$) due to the higher temperatures reach inside the star (see Fig. \ref{hrd}).

For the 0.4 $M_{\odot}$ star, the metal-poor model possesses the highest $\alpha$ and $\beta$ parameters because of its more extended and more massive radiative core. 
However, the evolution tracks of the mass and radius aspect ratios for the Z=0.0134 and Z=0.0255 stars are very similar and are almost superimposed which suggests that metallicity has only a minor effect on the internal structure of metal-rich low-mass star. 
Note here that the difference in metallicity between the Z=0.0134 and Z=0.0255 stars is lower than between the Z=0.004 and Z=0.0134 stars, so this conclusion should be verified by considering stars with an equal difference in metallicity.
For the 0.4 $M_{\odot}$ star the explanation of this behavior is the same as for the 1 $M_{\odot}$ star. 
For instance, the radiative core of 0.4 $M_{\odot}$ star starts to develop at 8.3, 15.1, and 17.8 Myr for the Z=0.004, Z=0.0134, and Z=0.0255 cases, respectively.

\begin{figure*}
\begin{center}
\includegraphics[angle=-90,width=0.4\textwidth]{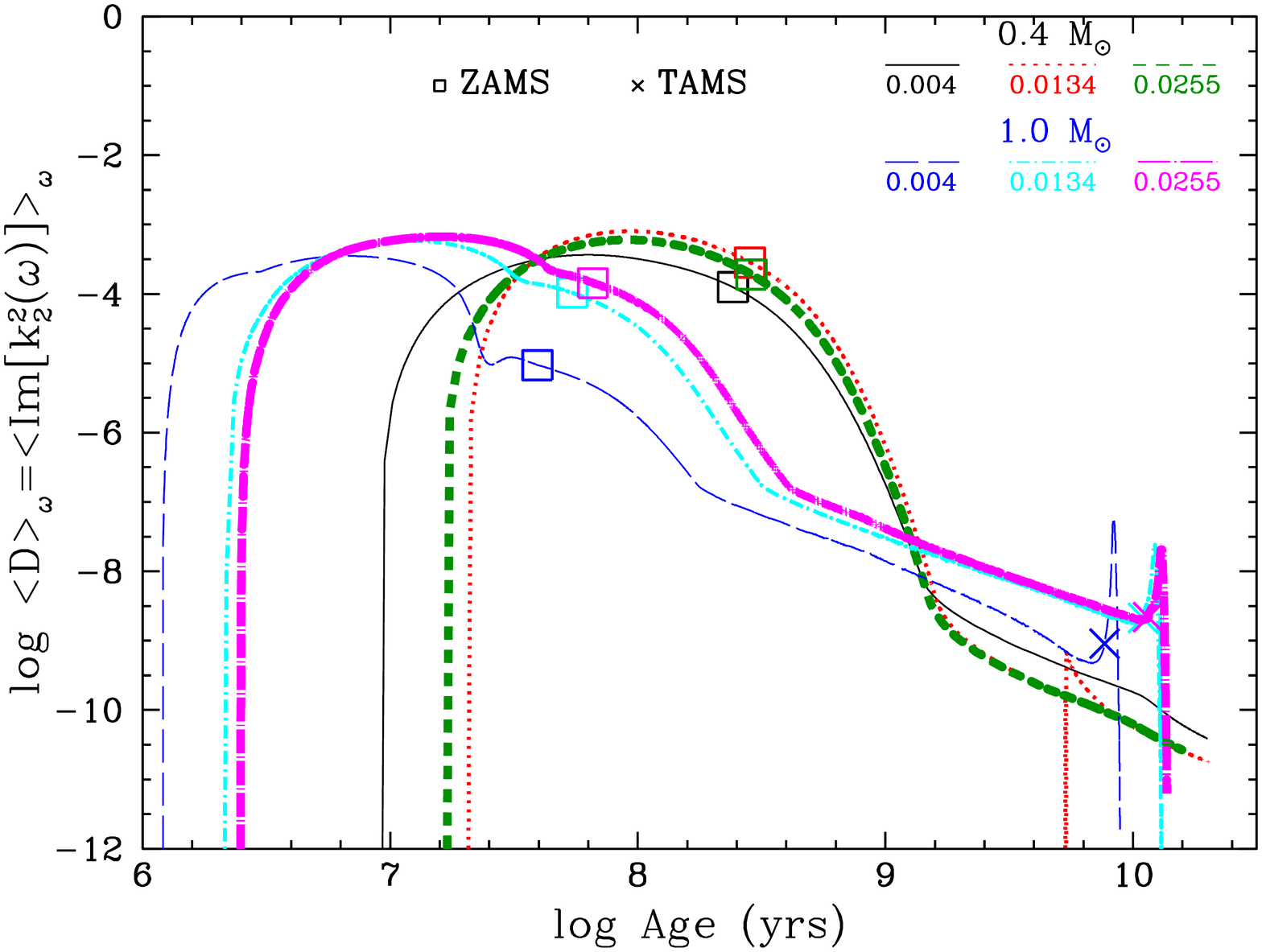}  \hspace{0.1cm} \includegraphics[angle=-90,width=0.4\textwidth]{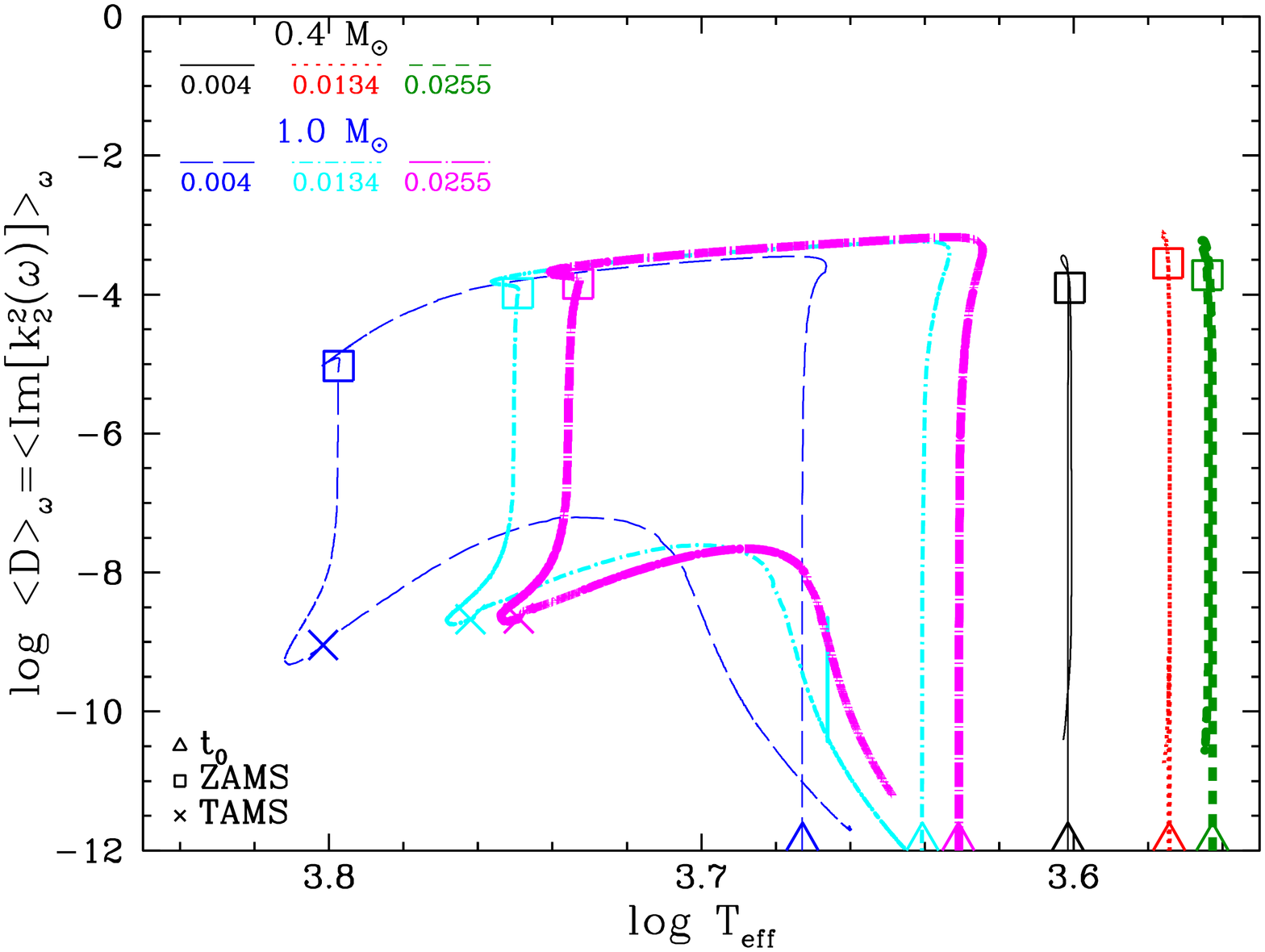}
\end{center}
\caption[bla]{Frequency-averaged tidal dissipation $<\mathcal{\hat{D}}>_{\omega}$ as a function of time (Left panel) and effective temperature (Right panel). The symbol represent: the first step in each model (triangle), ZAMS (square), and TAMS (cross)\protect\footnotemark.}
\label{Dissipz}
\end{figure*}

\subsection{Consequence for the dissipation}

Equation \ref{dissipequa} shows that the frequency-averaged tidal dissipation primarily depends on the internal structure of the stars (see Paper I for details).
For a given stellar mass, the evolution of the internal structure depends on the initial metallicity.

Figure~\ref{chemin} shows as a color coded gradient the intensity of the frequency-averaged tidal dissipation at fixed normalized angular velocity ($<\mathcal{D}>_{\omega}^{\Omega}$, see Eq. 3 in \citealt{Mathis15}) as a function of both radius and mass aspect ratios ($\alpha$ and $\beta$ respectively). 
The right-hand lower white region of Fig.~\ref{chemin} does not correspond to a high tidal dissipation intensity but to a non physical ($\alpha$, $\beta$) area where the condition $\gamma < 1$ is not fulfilled, meaning that it corresponds to a envelope denser than the core. 
Fig.~\ref{chemin} simultaneously displays the evolution of the couple ($\alpha$, $\beta$) along the evolution of stars of 0.4 and 1.0 $M_{\odot}$ models for the three metallicities considered here.

As stars evolve, they move towards higher values in the ($\alpha$, $\beta$) parameter space and pass close to the islet of maximum intensity ($\alpha_{\rm{max}}=0.572, \beta_{\rm{max}}=0.503$) corresponding to an intensity of $<\mathcal{D}>_{\omega}^{\Omega} = 1.091\times10^{-2}$ during the PMS, after which $\alpha$ and $\beta$ remain quite constant. 
When the stars are almost fully convective, they are in a regime that corresponds to regular inertial waves for which dissipation is weak \citep{Wu05}, conversely strong dissipation can be produced when the size and mass of the radiative core is sufficiently important to form sheared wave attractors \citep{Ogilvie07}.
Finally after the TAMS, the stars move in direction of lower values in the ($\alpha$, $\beta$) parameter space while they evolve towards the RGB phase.

The effect of a change of metallicity is clearly visible in the case of the metal-poor 1.0~$M_{\odot}$ star for which the amplitude between the maximum intensity (end of PMS phase) and the minimum intensity (MS-RGB phases) is higher than the metal-rich cases.
For the 0.4 $M_{\odot}$ models, the evolutions of $\alpha$ and $\beta$ are quite similar between the three metallicities. 
The metal-poor model eventually reaches higher intensity values in the ($\alpha$, $\beta$) parameter space during the MS, and thus higher dissipation.

Figure~\ref{Dissipz} shows the evolution of the frequency-averaged tidal dissipation (see Eq.~\ref{dissipequa}), for the three metallicities considered, as a function of time (left panel) and effective temperature (right panel).

Whatever the stellar mass, the general trends which appear when considering different metallicities remain the same.
The following description is therefore valid for the $0.4~\Msun$ star and the $1~\Msun$ star.
Following the evolution of the star inside the ($\alpha$, $\beta$) parameter space, the frequency-averaged tidal dissipation globally decreases when the metallicity is reduced modulo the evolution of the surface rotation of the star. 
This decrease can reach about two orders of magnitude at the ZAMS between the solar metallicity case (Z = 0.0134) and the metal-poor case (Z = 0.004).
For the metal-poor stars, we first observe that the dissipation is stronger than for metal-rich stars. 
This is mainly due to the delay of the apparition of the radiative core for the higher metallicities (see Fig.~\ref{stellarstructure}). 
Indeed, in metal-poor stars the apparition of the radiative core is linked to the metallicity (as metallicity decreases the averaged stellar opacity decreases and the stellar temperature increases). 
Hence, the tidal dissipation in the stars will start to increase first in metal-poor stars and then in stars more metal rich.  
As the size and mass of the radiative core reaches (for each metallicity cases) its nominal value, the observed trend of increasing dissipation with decreasing metallicity slowly reverses and the opposite behavior is then observed, namely the dissipation increases with metallicity.

\footnotetext{The grids of the frequency-averaged tidal dissipation and equivalent modified quality factor that we used in this work are available in an online tool: \url{https://obswww.unige.ch/Recherche/evol/starevol/Bolmontetal17.php.}}

\section{Effect of the metallicity of the star on the orbital evolution of close-in planets}
\label{orbits}

Since the dependence of the frequency-averaged tidal dissipation on the metallicity is quite significant, at least for the $1~M_\odot$ star, we explore here the consequences of those differences on the tidal evolution of close-in Jupiter-mass planets.
We choose here to consider the same initial time (i.e. the dissipation timescale of the protoplanetary disk) for all our simulations independently of the metallicity of the star or the initial rotation period\footnote{Note that the latter assumption is not compatible with the findings of \citealt{GB13} which showed that the disk lifetime depends on the initial rotation rate of the star. However, this is out of the scope of this paper and will be correctly taken into account later on.}. 
This allows us to isolate the effect of  metallicity on the orbital evolution of planets. 
We use the tidal evolution model presented in \citet{Bolmont16} and explored in Paper I for the case of solar-metallicity stars.

\subsection{Planets around fast-rotating metal-rich stars migrate farther away}
\label{fastrotatingstars}

\begin{figure}
\begin{center}
\hspace{+0.1cm}
\includegraphics[width=\linewidth]{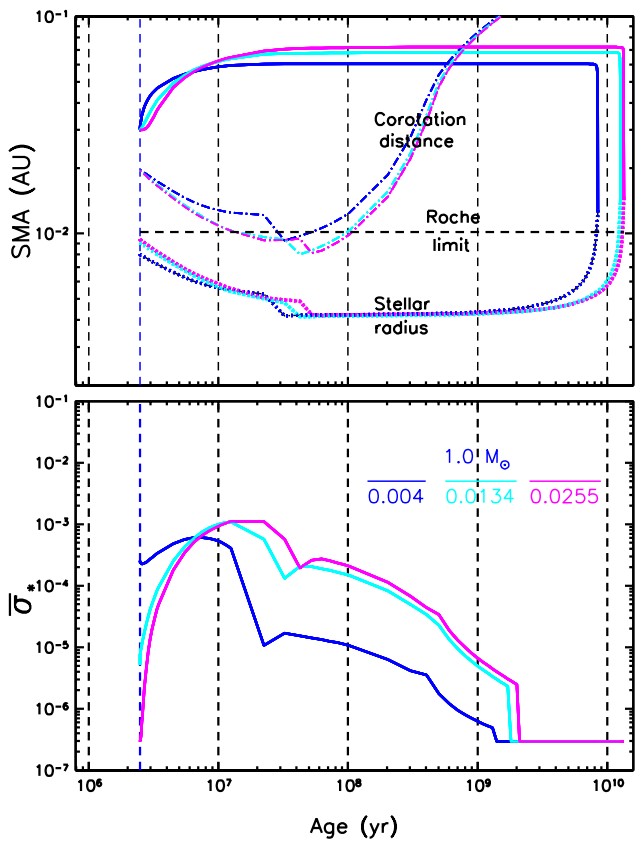}
\end{center}
\caption{\textbf{Top}: Evolution of the semi-major axis (full lines) of a close-in Jupiter mass planet around initially fast rotating $1~M_\odot$ stars of different metallicities: Z = 0.004 in dark-blue, Z = 0.0134 in cyan and Z = 0.0255 in pink. The dash-dotted lines correspond to the corotation radius. The colored dotted lines correspond to the radius of the star and the black dashed line corresponds to the Roche limit. \textbf{Bottom}: Evolution of the dissipation factor of the different $1~\Msun$ stars.}
\label{sma_diff_Z}
\end{figure}

Figure~\ref{sma_diff_Z} shows the evolution of the semi-major axis of a Jupiter mass planet orbiting a 1~$\Msun$ star of three different metallicities (Z = 0.004, 0.0134, and 0.0255).
The planet tidally evolves from the moment of disk dispersal, which is here considered to be 2.5~Myr for all the models.
We consider fast stellar rotators (initial $P_{\star, \rm{0}} = 1~$day) and assume that the planets begin their evolution at a semi-major axis of 0.03~au. 
Fig.~\ref{sma_diff_Z} also shows the evolution of the corotation radius, which is the orbital distance at which the planet's orbital period is equal to the stellar surface rotation period (i.e., where $n = \Omega_\star$). 
It separates the region in which planets migrate inwards from the region in which planets migrate outwards.
The corotation radius therefore evolves as the star spins up for the first few $10^7$~yr of evolution and spins down afterwards due to the stellar wind \citep[see][]{Bolmont11,Bolmont12,Bolmont15,Bolmont16}.

\captionsetup[figure]{position=top}
\begin{figure*}
\centering
\subcaptionbox{$P_{\star,0} =$ 1 day \label{model1days}}
{\includegraphics[width=0.35\linewidth]{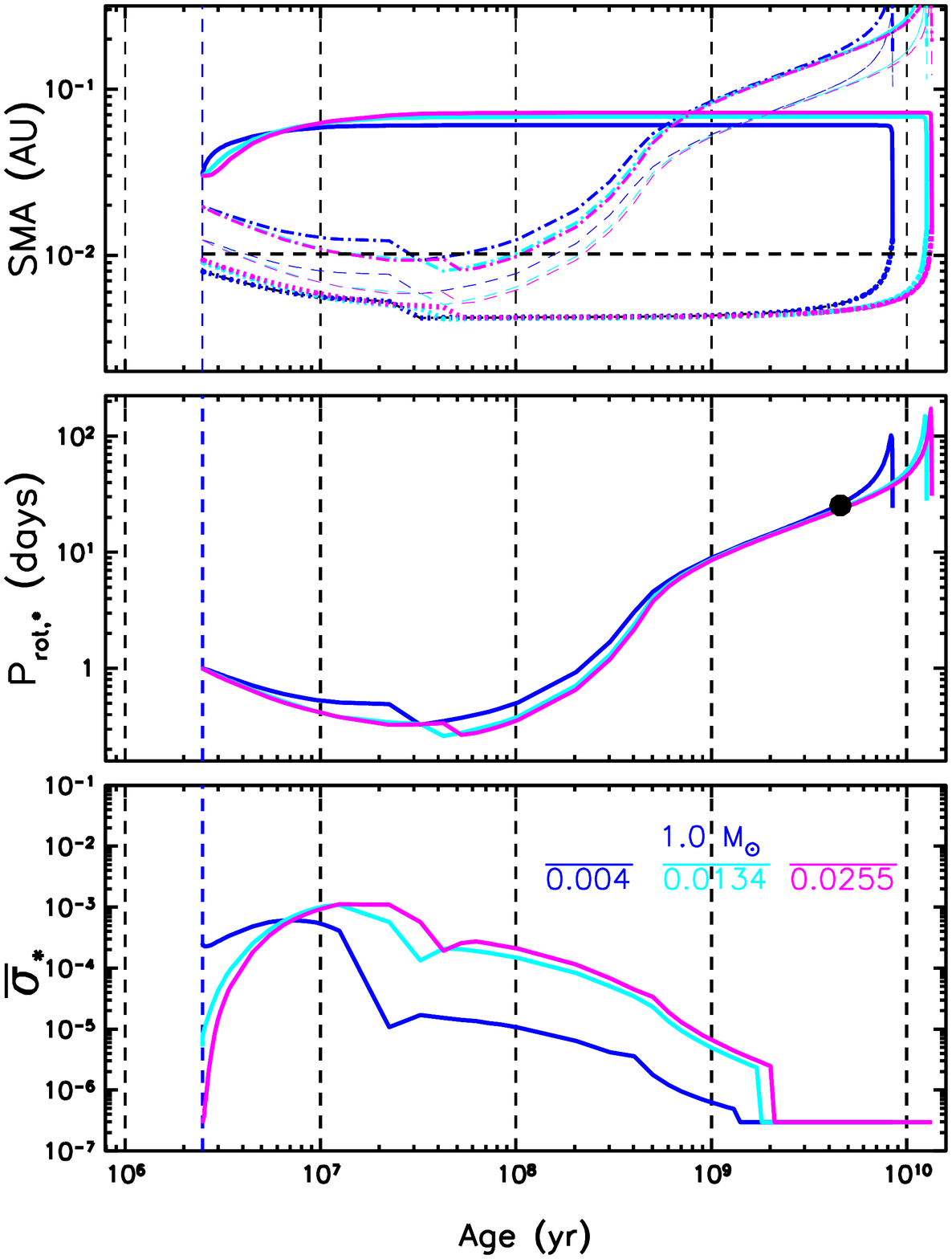}
}\hspace{-0.9cm}
\subcaptionbox{$P_{\star,0} =$ 3 day \label{model3days}}
{\includegraphics[width=0.35\linewidth]{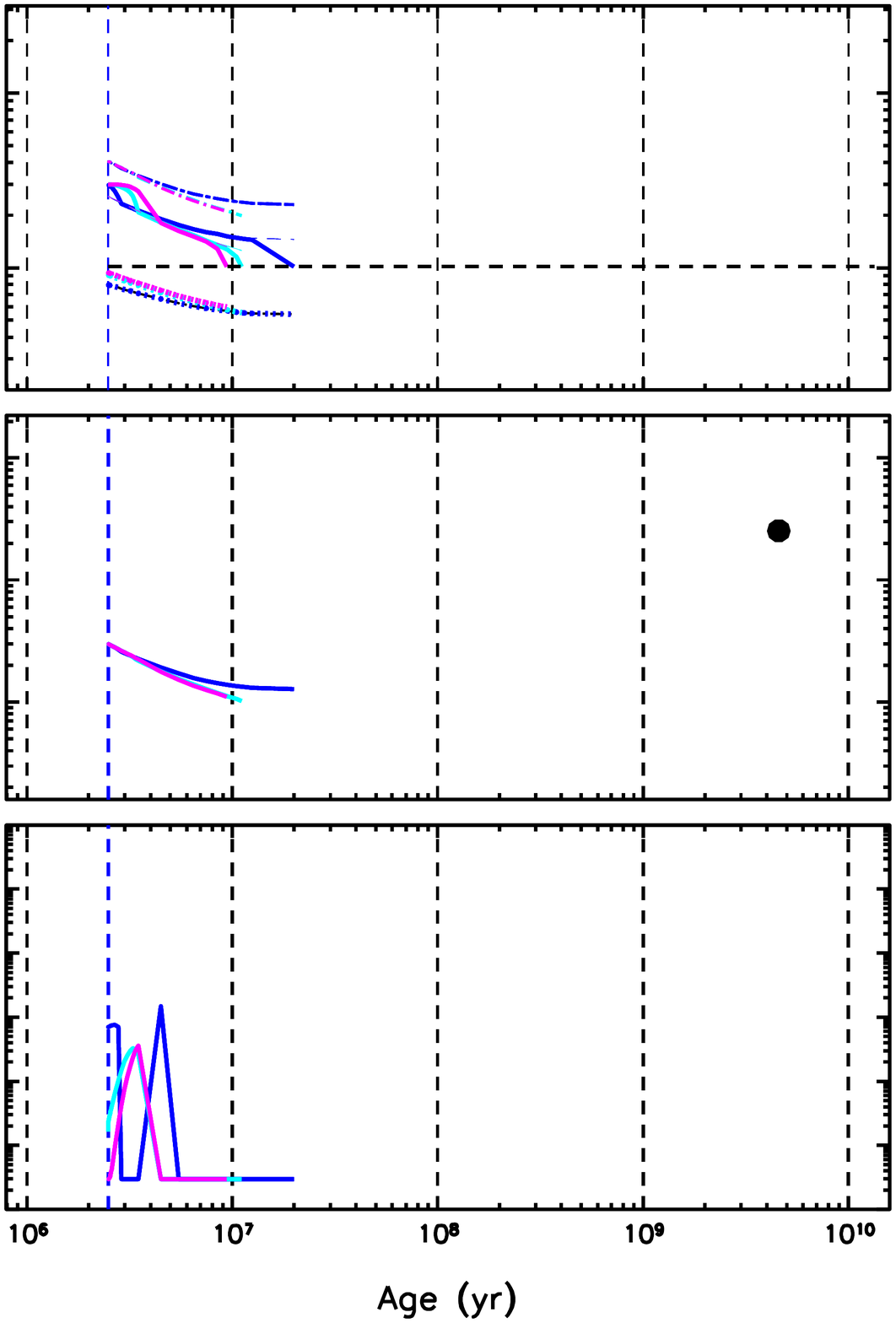}
}\hspace{-0.9cm}
\subcaptionbox{$P_{\star,0} =$ 8 day \label{model8days}}
{\includegraphics[width=0.35\linewidth]{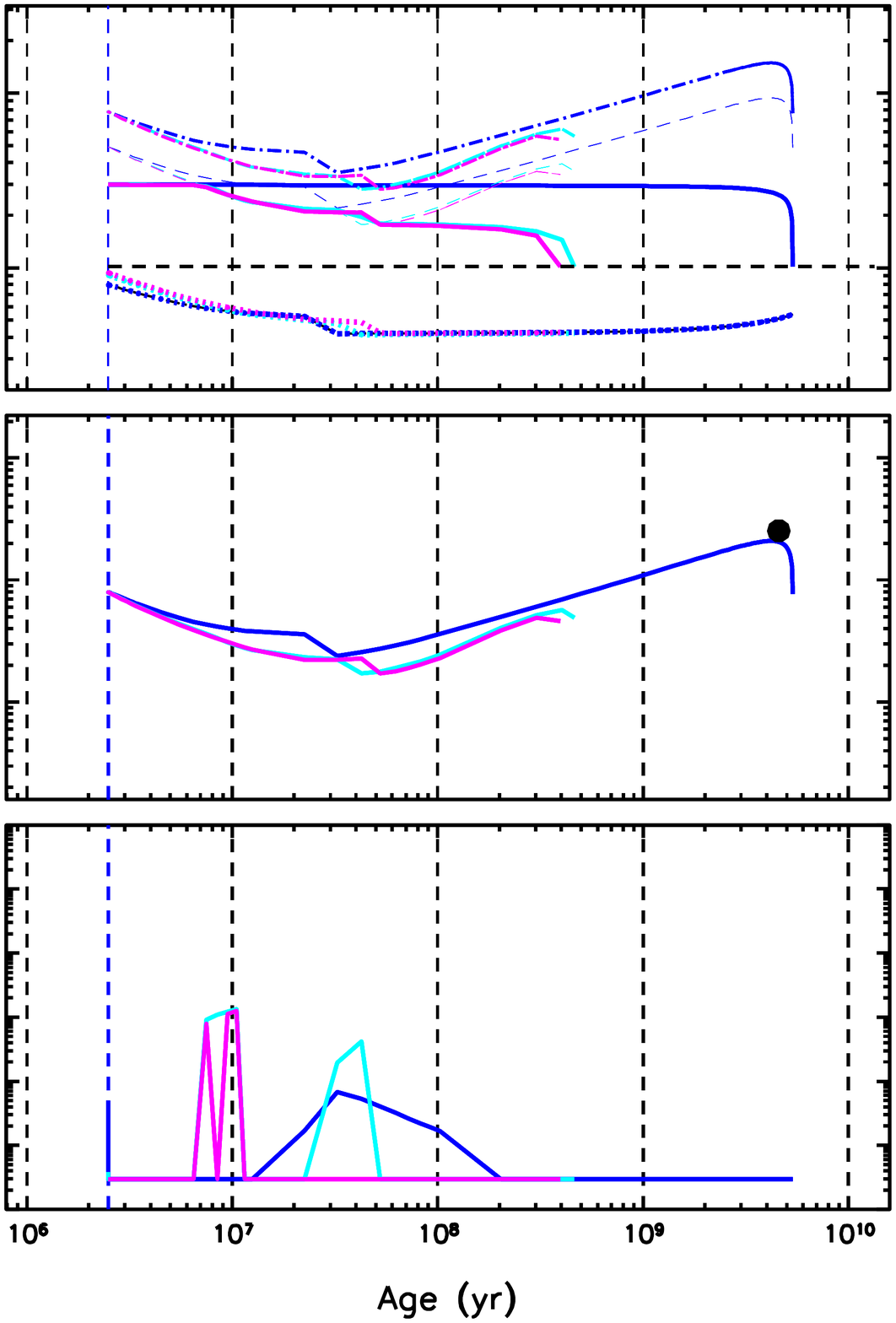}
}
\caption{Tidal evolution of a close-in Jupiter mass planet around stars of different metallicities (Z = 0.004 in purple, Z = 0.0134 in dark blue and Z = 0.0255 in light blue) and different initial rotation periods: a) $P_{\star, 0} = 1~$day, b) $P_{\star, 0} = 3~$day and c) $P_{\star, 0} = 8~$day. \textbf{Top}: Evolution the semi-major axis (full lines), corotation radius (dashed-dotted lines), line defining $P_{\rm{orb}} = 1/2~P_\star$ (thin long dashes), stellar radius (dotted lines), the Roche limit (black dashed line). \textbf{Middle}: stellar rotation period evolution, the black dot represents the rotation period of the Sun today. \textbf{Bottom}: Evolution of the dissipation factor for the different stars.}\label{sma_diss_diff_Z_diff_P0}
\end{figure*}

Initially, the planets we consider are outside the corotation radius and migrate outwards.
Fig.~\ref{sma_diff_Z} shows that during the first million years of evolution, the planet orbiting the metal-poor star migrates faster.
Only considering the upper panel of Fig.~\ref{sma_diff_Z}, this might appear counter-intuitive because the radius of the metal-poor model is initially lower than the radius of the other ones.
As Eq.~\ref{Ts} shows, the lower the radius, the higher the evolution timescale (all other things being equal).
This difference comes from the fact that the dissipation (see Fig.~\ref{Dissipz}) and the dissipation factor (see lower panel of Fig.~\ref{sma_diff_Z}) of metal-poor stars are initially much higher than those of metal-rich stars.
Fig.~\ref{sma_diff_Z} actually shows that the dissipation factor of the Z = 0.004 star is initially more than one order of magnitude higher than that of the Z = 0.0134 star and three orders of magnitude higher than that of the Z = 0.0255 star.
This higher dissipation does more than compensate for the lower radius of the metal-poor star and leads to an initial faster outwards migration.

As the 1~$\Msun$ stars evolve, the dissipation factor increases whatever the metallicity, and around $\sim 6$~Myr, the dissipation factor of the two metal-rich stars becomes larger than the dissipation factor of the metal-poor star.
From that moment on and until the end of the MS, the radius of the two metal-rich stars is larger than that of the metal-poor star and the dissipation factor is bigger, which leads to a faster migration for the corresponding planets.
Fig.~\ref{sma_diff_Z} shows that after a few Myr, the planets around the metal-rich stars migrate farther away and during a longer time than the planet around the metal-poor star.
As a consequence, we can say that \textit{the planets around initially fast-rotating metal-rich stars should be located farther away than planets around initially fast-rotating metal-poor stars}.
For example, at the age of $5~$Gyr, the planets around the Z = 0.004, 0.0134, and 0.0255 solar mass models migrated from 0.03~au to 0.061~au, 0.068~au, and 0.072~au respectively.
This is more than 0.01~au difference between the planets orbiting the metal-poor star and the metal-rich star, which is not negligible.

The planets cross the expanding corotation radius around $600$ to $800$~Myr, and proceed to very slowly migrate inwards (with timescales as long as 10~Gyr).
Indeed, as they cross the limit $P_{\rm orb} = 1/2~P_\star$, the tidal inertial waves are no longer excited in the convective region and the dissipation factor falls to the equilibrium tide value ($\overline{\sigma_\star} = 3\times 10^{-7}$, see Fig.~\ref{sma_diff_Z} and \citealt{Bolmont16}). 

As the star evolves towards the end of the MS and up the RGB phase, the radius starts increasing again.
As the planet orbiting the metal-poor star is closer and the stellar radius is increasing faster than for the two other stars, the planet falls onto the star earlier in its history, at 8.4~Gyr.
Note that the planet is here actually engulfed in the star before being tidally disrupted.
The planet therefore spirals inwards in the upper layers of the star before reaching the Roche limit and being destroyed.
The planets orbiting the most metal-rich stars fall onto the star later in the evolution, around an age of $13$~Gyr.
This is due to the fact that they are farther away and that the radii of the stars increase on longer timescales, postponing the engulfment event.

\medskip

We also investigated the influence of the metallicity for the $0.4~\Msun$ star.
The trend we discussed is much less visible than in the case of a $1~\Msun$ star.
For example, for planets beginning at 0.03~au around initially fast rotating stars, we find that at an age of 5~Gyr, they are at 0.035~au, 0.036~au and 0.037~au around the Z = 0.004, Z = 0.0134 and Z = 0.0255 stars respectively.

\subsection{Effect of the initial stellar spin}

In the previous section, we considered initially fast-rotating 1~$\Msun$ star. 
We consider here models for the same initial mass but with slower initial rotation (periods of 3 and 8~days).

Figure~\ref{sma_diss_diff_Z_diff_P0} shows the different evolutions for different initial spins.
For close-in planets ($a_{\rm init} = 0.03~$au), changing the initial stellar spin leads to very different evolutions.
For a fast rotating star, the planet survives and migrates outwards. 
While for slower rotating stars, the planet is initially inside the corotation radius and migrates inwards until it falls onto the star, or rather until it gets tidally disrupted when reaching the Roche limit.

\begin{sidewaysfigure*}
    \centering
\includegraphics[angle=0, width=\linewidth]{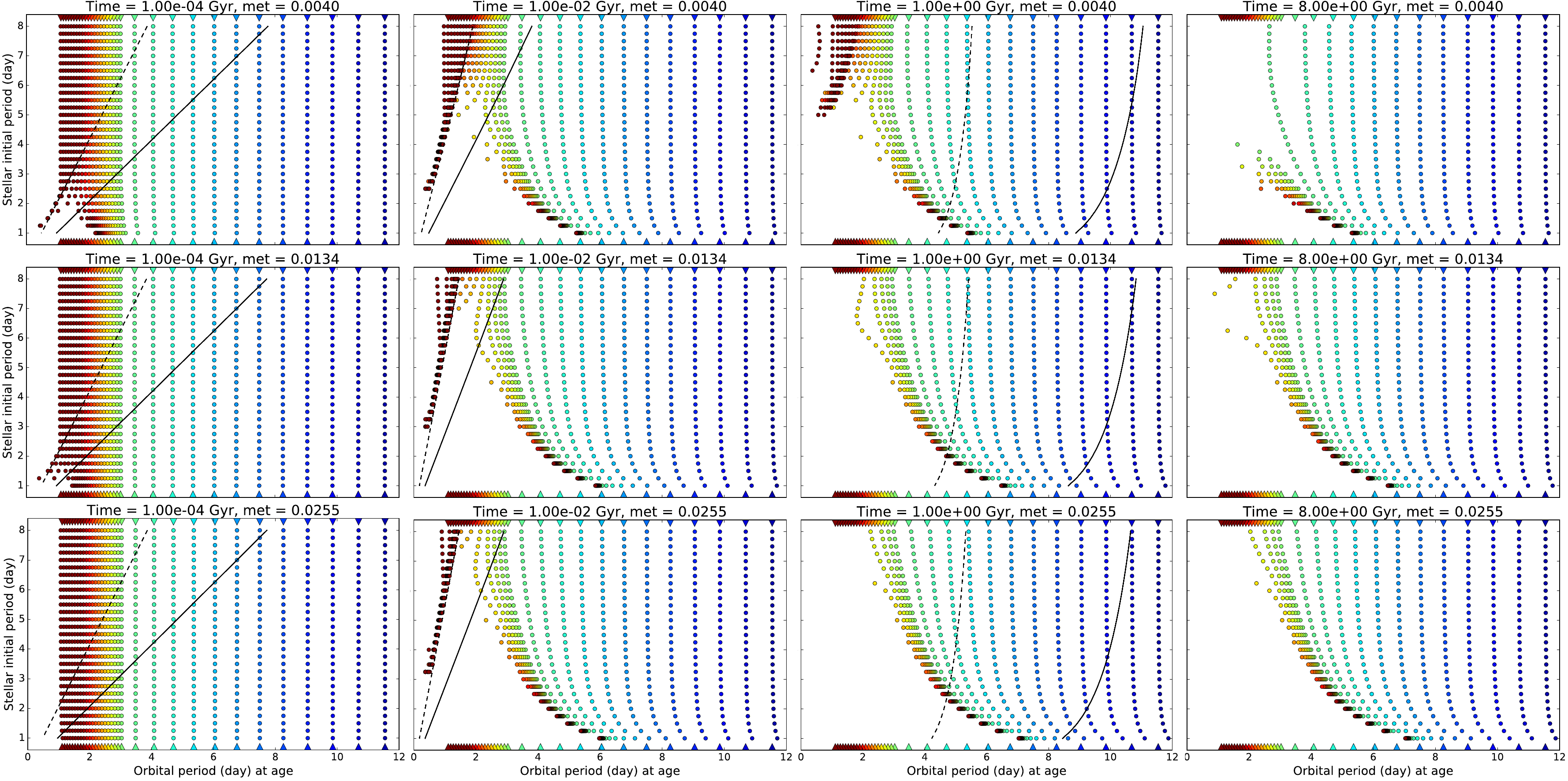}
\caption{Evolution of a population of hot Jupiters at different initial orbital periods around different initial rotation period of the star. The different panels are for a different time (time = Age - t$_{\rm init}$, increases from \textbf{left to right}) in the evolution of the system and different metallicities of the star (Z = 0.004, Z = 0.0134 and Z = 0.0255 from \textbf{top to bottom}). The position of each colored circles represent the orbital period of one planet (x-axis) at the time considered and the initial rotation period of the star it orbits (y-axis). The planets orbiting initially fast rotating stars are in the lower part of the diagram while the planets orbiting initially slow rotating stars are in the upper part. The color of the circle is an indication of the initial orbital period of the planet, the color-initial orbital period relationship is given by the colored triangles on the top and bottom of each panel. The dark red planets are initially close-in planets. The full black line represents the corotation radius (delimiting inside and outside tidal migration) and the dashed black line represents the limit between the equilibrium and dynamical tide regimes corresponding to the simulation of the planet at 0.1~au.}
\label{Pop_Age_infl_Z}
\end{sidewaysfigure*}

%\begin{figure*}
%\centering
%%\begin{center}
%%\includegraphics[angle=-90, width=10cm]{Pop_Age_infl_Z_2.pdf}
%\includegraphics[angle=-90, width=10.5cm]{Pop_Age_infl_Z_2_bigger.pdf}
%%\end{center}
%\caption{Evolution of a population of hot Jupiters at different initial orbital periods around different initial rotation period of the star. The different panels are for a different time (time = Age - t$_{\rm init}$, increases from \textbf{top to bottom}) in the evolution of the system and different metallicities of the star (Z = 0.004, Z = 0.0134 and Z = 0.0255 from \textbf{right to left}). The position of each colored circles represent the orbital period of one planet (x-axis) at the time considered and the initial rotation period of the star it orbits (y-axis). The planets orbiting initially fast rotating stars are in the lower part of the diagram while the planets orbiting initially slow rotating stars are in the upper part. The color of the circle is an indication of the initial orbital period of the planet, the color-initial orbital period relationship is given by the colored triangles on the top and bottom of each panel. The dark red planets are initially close-in planets. The full black line represents the corotation radius (delimiting inside and outside tidal migration) and the dashed black line represents the limit between the equilibrium and dynamical tide regimes corresponding to the simulation of the planet at 0.1~au.}
%\label{Pop_Age_infl_Z}
%\end{figure*}

Figure~\ref{model3days} shows the planet's evolution for a stellar initial period of 3~days.
Initially the planets are inside the corotation radius but above the line corresponding to $P_{\rm orb} = 1/2~P_\star$, meaning that they migrate inwards due to the dynamical tide.
All planets migrate inwards rapidly until reaching the $P_{\rm orb} = 1/2~P_\star$ line. 
As discussed in \citet{Bolmont16}, once the planet reaches this line, the equilibrium tide takes over.
However, as the corresponding dissipation factor is much lower, the inward migration slows down. 
Meanwhile the star continues shrinking and spinning up, leading to the decrease of the $P_{\rm orb} = 1/2~P_\star$ limit.
The planet enters once again the region of excitation of the inertial waves and the dissipation factor increases to the dynamical tide value (peaks in the last panel of Fig.~\ref{model3days} and \ref{model8days}). 
The inward migration speeds up again due to the higher dissipation factor until reaching once again the $P_{\rm orb} = 1/2~P_\star$ limit. 
This capture process keeps on until the planet is close enough to fall onto the star rapidly due to the equilibrium tide.
One interesting observation is that, even though the planet around the Z = 0.004 star migrates inwards faster during the first few million years, it finally falls last onto the star.
This is due to the fact that for a given disk lifetime, higher metallicity stars accelerate more on the PMS phase than metal-poor stars (see Fig.~\ref{sma_diff_Z}), so that the capture of the planets on the $P_{\rm orb} = 1/2~P_\star$ limit occurs for smaller orbital distances (see Fig.~\ref{model3days} or for an earlier age (see Fig.~\ref{model8days}).

Figure~\ref{model8days} shows the planet's evolution for a stellar initial period of 8~days.
Initially the planets are inside the corotation radius and below the line corresponding to $P_{\rm orb} = 1/2~P_\star$, meaning that they migrate slowly inwards due to the equilibrium tide.
Meanwhile the star spins up and after a few million years, the planet reaches the dynamical tide region.
It reaches it earlier for more metal-rich stars as they spin up faster than the metal-poor stars.
The capture mechanism previously described occurs and planets migrate inwards as the $P_{\rm orb} = 1/2~P_\star$ limit decreases until an age of $\sim 50$~Myr.
They then migrate slowly inwards due to the equilibrium tide and get disrupted reaching the Roche limit at an age of 300-400~Myr.
However, the planet around the metal-poor star reaches the dynamical tide region later, at an age of about $\sim 10$~Myr. 
At that age, the dissipation factor of this star is already lower than that of the two metal rich stars.
The planet is captured for a while on the $P_{\rm orb} = 1/2~P_\star$ limit while the star experiences a slow evolution (in radius and spin). 
At an age of 20~Myr, the star spins-up faster due to the rapid decrease of the radius (see Figs.~\ref{stellarstructure} and \ref{sma_diff_Z} for a zoomed view). 
The dynamical tide induced inward migration is not fast enough to ensure that the planet stays captured on the $P_{\rm orb} = 1/2~P_\star$ limit so that the planet is back into the dynamical tide region.
However, at that age, the dynamical tide dissipation factor is already quite low and the planet does not experience a fast inward migration.
It then leaves the dynamical tide region and proceed to spiral inwards on Gyr timescales to be tidally disrupted at an age of $\sim 5~$Gyr.
To conclude, \textit{very close-in planets around metal-poor stars with an initial rotation from moderate to slow survive longer than planets around metal-rich stars.}

\subsection{Effect of the metallicity on a population of Hot Jupiters}

In this section, we investigate the effect of the metallicity on the evolution of an initial population of hot Jupiters.
We consider planets with an initial orbital period between 1.03 day (0.02~au) and 11.5 day (0.1~au) orbiting stars with an initial rotation period between 1 and 8 days.
Figure~\ref{Pop_Age_infl_Z} shows the evolution of this population for the three different metallicities (Z = 0.004, Z = 0.0134 and Z = 0.0255) with four snapshots: $10^{-4}$~Gyr ($10^5$~yr, almost representative of the initial distribution), 0.01, 1.0 and 8.0~Gyr after the beginning of the simulation.

If we consider the first column corresponding to an early time of 100,000 yr after the beginning of the simulation (100,000 yr after the protoplanetary disk dispersal), we can see that the evolution around the metal poor star is occurring faster than around the two other stars.
The area just above the corotation radius of the fast rotating stars is emptied quite efficiently pushing outwards the planets outside corotation, and inwards the planets inside.
The area between the corotation radius and the limit equilibrium-dynamical tide is also emptied for the fast rotating stars: the planets migrate quickly to the limit $P_{\rm orb} = 1/2~P_\star$ where they pile up for a while. 
As discussed above, this pile-up comes from the fact that the equilibrium tide dissipation is much lower than the dynamical tide dissipation.
This phenomenon is starting around the Z = 0.0134 stars but nothing is yet visible around the Z = 0.0255 stars.

After 10~Myr of evolution (second column of Fig.~\ref{Pop_Age_infl_Z}), the population has changed quite dramatically already: the inner regions around initially fast rotating stars have been completely emptied.
If the initial stellar rotation period is less than 2.5 days, every planet on a close-in orbit around the Z = 0.004 star has either migrated outwards significantly or has collided with the star.
For the metal rich stars, this limit increases to 3 days.
Around the initially slow rotating stars, a population of close-in planets in the process of falling due to the equilibrium tide survives still.
We can observe the pile up of captured planets on the $P_{\rm orb} = 1/2~P_\star$ limit.

After 1~Gyr (third column of Fig.~\ref{Pop_Age_infl_Z}), the inner regions around the two metal rich stars are depleted, especially around the initially fast rotating stars.
Around the initially slow rotating stars, the population that was in the process of falling due to the equilibrium tide at 10 Myr has completely disappeared, only remains the population which had initial orbital periods greater than 2.5 days. 
The evolution of this population is due to the equilibrium tide which drives the inward migration on a few gigayear timescale. 
This is illustrated by the small differences between the 1~Gyr and the 8~Gyr snapshots (lower two panels, last two columns of Fig.~\ref{Pop_Age_infl_Z}).
This close-in population of planets around initially slow rotating stars, that was impacted by the dynamical tide in its early history, disappears in about 0.5~Gyr for the Z = 0.0255 star, in about 1~Gyr for the Z = 0.0134 star and in about 5.5~Gyr for the Z = 0.004 star. 

While the tidal evolution around the two metal rich stars is not very pronounced after 1~Gyr of evolution (all the planets either fell onto the star or migrated outwards sufficiently not to feel the tides), migration around the metal poor star occurs on longer timescales: 
\begin{enumerate}
\item[--] The population of initially very close-in planets (dark red planets in Fig.~\ref{Pop_Age_infl_Z}, first line, third panel) around the initially slow rotating stars is still in the process of falling due to the equilibrium tide at that time;
\item[--] The population of planets around the stars with an initial rotation period of 2 to 4 days has not migrated outwards sufficiently to stop being influenced by tides and are slowly spiraling inwards.
\end{enumerate}

Figure~\ref{Pop_Age_infl_Z} shows an interesting fact among the population of planets around slowly rotating metal poor stars at 1~Gyr (last column, third raw), more specifically for stars with an initial rotation period between 5 and 6~day.
Planets in orange and light red (corresponding to planets initially with orbital period between 2 and 2.5~day) can be found closer-in than planets in dark red (corresponding to planets initially with orbital period smaller than 2~day).
In other words, the planets orbiting a metal poor star initially slowly rotating with an initial orbital period from 2 to 2.5 days collide with the star before the very close-in planets ($P_{\rm orb} \sim 1$~day) do.
This appears counter-intuitive because planets initially farther away should migrate on longer timescales. 
Once again, this is due to the interplay between the dynamical tide and equilibrium tide.
This phenomenon is not visible for initially fast rotating stars for which almost all planets migrate outwards and survive the orbital evolution.

\begin{figure}
\begin{center}
\includegraphics[width=0.45\textwidth]{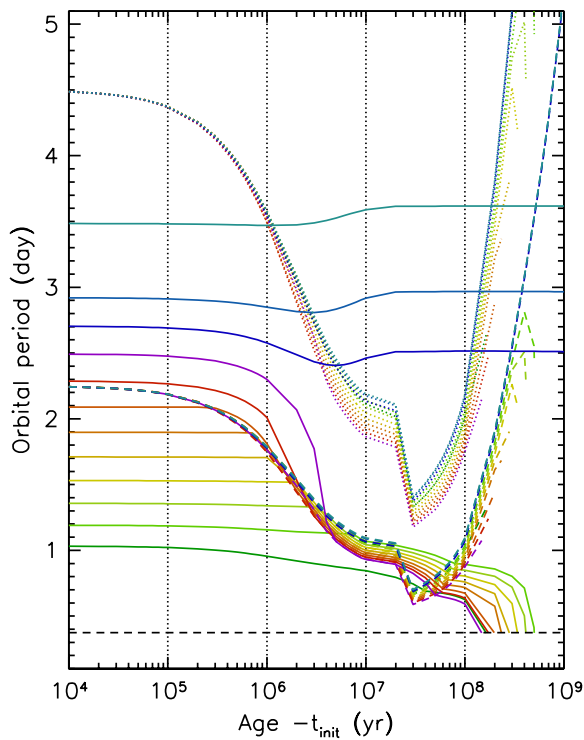}
\end{center}
\caption{Evolution of close-in Jupiter mass planets around a $1.0~\Msun$ star of metallicity Z = 0.004 with an initial rotation period of 4.5~day. The coloured full lines correspond to the orbital period evolution of the planets. The colored dashed-dotted lines correspond to the rotation period of the star and the colored dashed lines correspond to the $1/2~P_\star$ limit. The thick black dashed line corresponds to the Roche limit.}
\label{mid_sma_crash_before_small_sma}
\end{figure}

Figure~\ref{mid_sma_crash_before_small_sma} shows the evolution of this population for a Z = 0.004 star with an initial spin of 4.5~day. 
For this case, all planets with initial orbital periods smaller or equal to 2.6~days fall onto the star before 1~Gyr, while planets initially farther away survive the first Gyr of evolution due to the spin up of the star.
For the planets initially closer than 2.6~day, the farther is the planet, the earlier it falls onto the star.
Let us take the example of the planet beginning at an orbital period of 2.5~day (full violet line in Fig.~\ref{mid_sma_crash_before_small_sma}). 
It is initially in the region where the dynamical tide drives the evolution, it therefore migrates inwards rapidly. 
The transfer of angular momentum between the planet's orbit and the stellar spin makes the star spin up significantly, so that the $P_{\rm orb} = 1/2~P_\star$ limit decreases with time. 
This planet is initially sufficiently close to be captured on the $P_{\rm orb} = 1/2~P_\star$ limit after about 5~Myr of evolution.
From this moment on and for 15~Myr, the planet stays captured on this $P_{\rm orb} = 1/2~P_\star$ limit and migrates inwards as the star spins up.
When we say that the planet is on the limit, in reality it moves regularly between the two regions (see \citealt{Bolmont16}) and when it is in the dynamical tide dominated regime, the sudden inward migration acts to spin up the star. 
At 20~Myr, the stellar spin increases rapidly due to a rapid decrease of the stellar radius (seen in Fig.~\ref{sma_diff_Z}). 
The planet is back in the region where the dynamical tide drives the evolution and leaves it after a few $10~$Myr when the contraction is over and the star begins to spin down due to the stellar wind.
It collides on the star after $\sim$165~Myr.
When a planet is initially closer-in, the inward migration occurring before the capture of the planet on the $P_{\rm orb} = 1/2~P_\star$ limit is less important than the previous case for two reasons:
\begin{enumerate}
\item[--] Either the planet is initially in the equilibrium tide region and it does not migrate inward significantly and gets captured when the star has spun-up enough so that $P_{\rm orb} = 1/2~P_\star$;
\item[--] Or the planet is initially in the dynamical tide region and migrates inward faster than the previous case thus catching the $P_{\rm orb} = 1/2~P_\star$ limit earlier in the evolution.
\end{enumerate}
The fate and collision time of the planets is decided in the first million years of evolution, depending on its initial position with respect to the $P_{\rm orb} = 1/2~P_\star$ limit.

\subsubsection*{Statistical trends: comparison with observations}

\begin{figure}
\begin{center}
\includegraphics[angle=-90,width=0.4\textwidth]{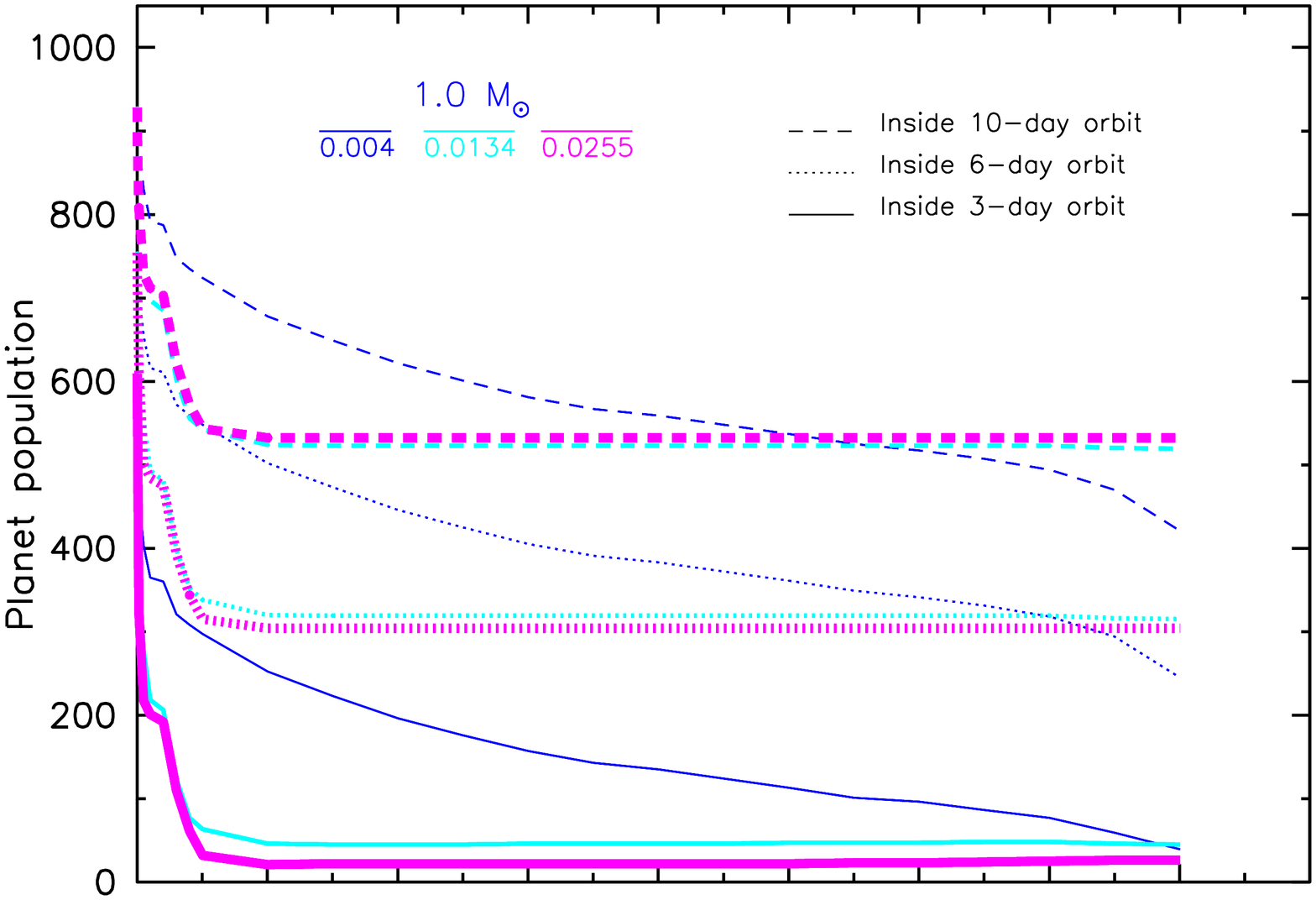}\\  \vspace{-0.70cm} \includegraphics[angle=-90,width=0.4\textwidth]{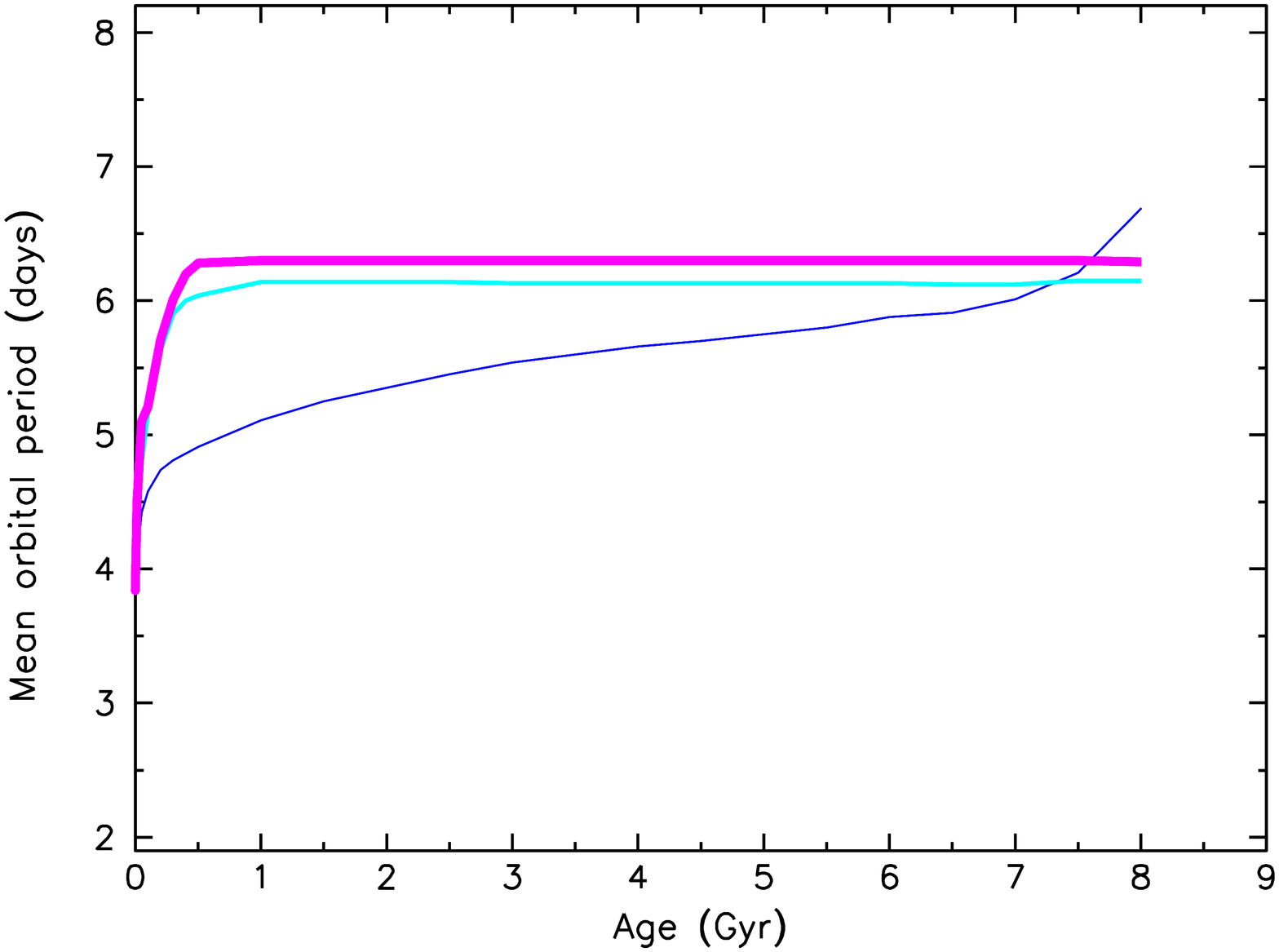}
\end{center}
\caption{\textbf{Top}: Evolution of the number of planets within a 3-day, 6-day and 10-day orbit for the three different metallicities considered here. \textbf{Bottom}: Evolution of the mean orbital period of the simulated planets for the three different metallicities.}
\label{number_inside_Pdays_legend}
\end{figure}

The conclusion we made in Section \ref{fastrotatingstars}, that planets should be found farther away from fast rotating stars does not appear to be generalizable to slower initial rotations. 
In order to quantify that, we monitored for each metallicities the population of planets interior to a given orbital period at a given age.
The top panel of Fig.~\ref{number_inside_Pdays_legend} shows differently the fact that the population of planets around metal poor stars changes during the whole stellar evolution, while the population of planets around the metal rich stars changes a lot at early ages and then stays constant during 7 Gyr.
The bottom panel of Fig.~\ref{number_inside_Pdays_legend} also shows how the mean orbital period of the planets varies with time.

Let us assume there is no particular prior distribution of planets after the formation phases depending on stellar initial periods and metallicities, and that the distribution of stars according to their initial rotation and metallicity period is uniform. 
In other words, we consider the same initial population for all stars and therefore do not take into account the fact that metal poor stars might form less hot Jupiters than metal rich stars or the fact that the formation location of the planets might depend on the metallicity.
With this hypothesis, the long-term evolution of the population of planets around the metal poor star leads to the following observations:
\begin{enumerate}
\item[--] Around young Sun-like stars (Age $< 5-6$~Gyr), there should be statistically more planets ($P_{\rm orb} < 10$~days) around metal poor stars than around metal rich stars, and those planets should be located closer;
\item[--] Around old Sun-like stars (Age $> 5-6$~Gyr), there should be statistically fewer planets ($P_{\rm orb} < 10$~days) around metal poor stars than around metal rich stars. For even later ages (Age $>7$~Gyr), those planets should be located farther;
\end{enumerate}
How do these conclusions about our simulations compare to the observations of statistical trends in the hot Jupiters population? 

\medskip

The relationship between presence of a Hot Jupiter and the metallicity of the star has been intensively studied over the years both on the theoretical side \citep{Mordasini2009,Mordasini12,Johnson2012,Hasegawa2014,Pinotti2017} and the observational side \citep{Gonzalez1997,Santos2003,Neves2013,Mortier2013,Adibekyan13,Mulders2016}.

\medskip

\begin{figure}
\begin{center}
\includegraphics[angle=0,width=0.5\textwidth]{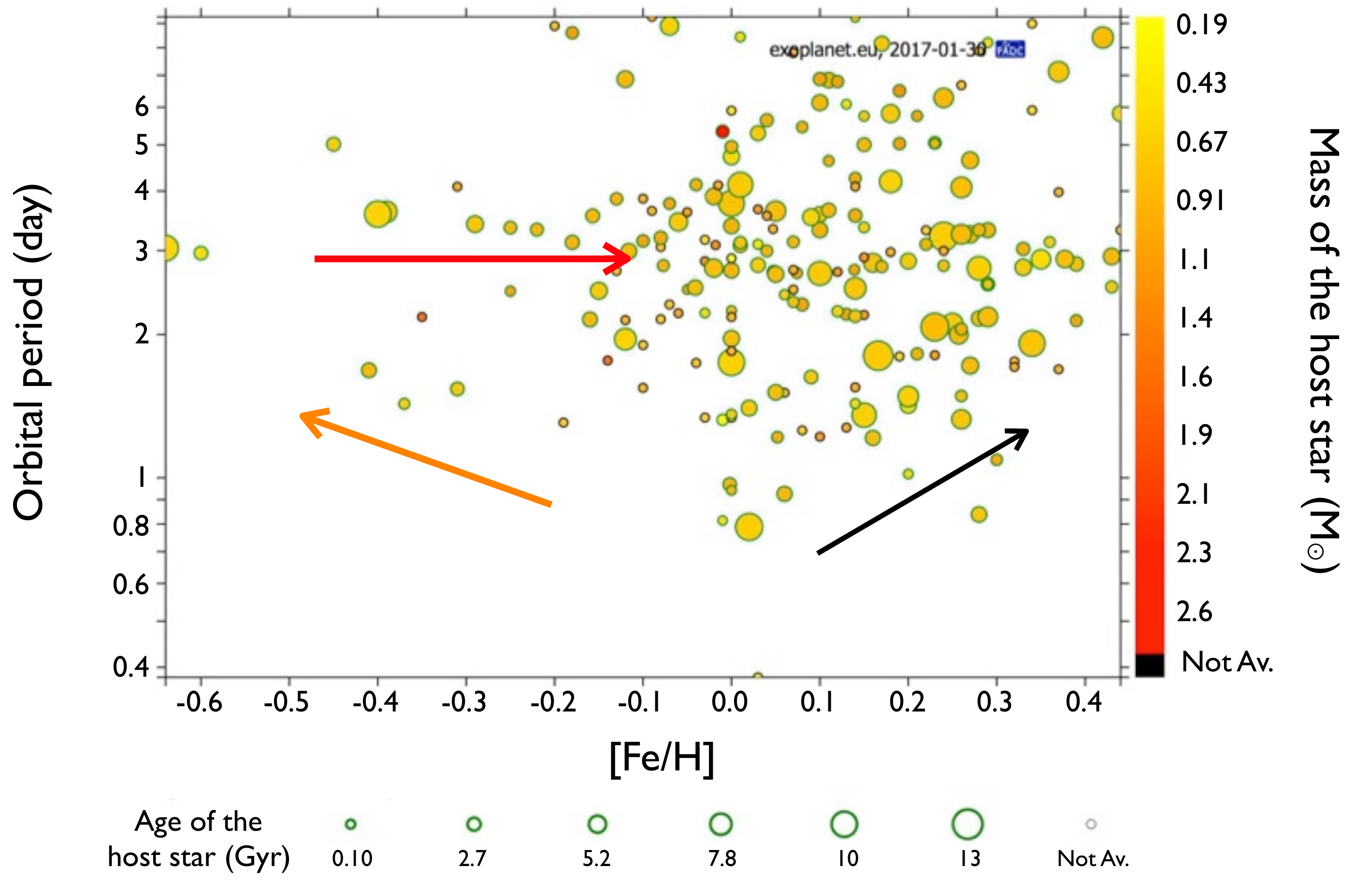}
\end{center}
\caption{Population of hot Jupiters with a mass higher than $0.8~M_{J}$ and a period lower than 10~day. Figure adapted from \url{http://exoplanet.eu/}. The arrows represent the statistical trends discussed in the text.}
\label{pop_exoplanet_eu_per_vs_met}
\end{figure}

On the one hand, the observations tend to show several trends. 
Figure~\ref{pop_exoplanet_eu_per_vs_met} shows the population of Hot Jupiters in a orbital period-metallicity diagram, which displays the two following trends:
\begin{enumerate}
\item[--] \textbf{Trend 1} (red arrow): the number of massive planets is higher around metal rich stars \citep{Neves2013}, 
\item[--] \textbf{Trend 2} (orange arrow): the massive planets around metal poor stars seem to be on wider orbits than the planets around metal rich stars \citep{Adibekyan13}.
\end{enumerate}
These trends could have three different origins: the first one is the eventual dependency of the initial planet population on the metallicity (see the discussion on theoretical planetary formation studies a few paragraphs down), the second one is the dependency of the planets' orbital evolution on the metallicity (what we are investigating at here) and the third one is an observational bias towards metal rich stars \citep[see discussion in][]{Santos2005}. 
We explore in the next two paragraphs the hypothesis that this distribution is solely due to the different tidal orbital evolutions the planets undergo around stars of different metallicities.

We find that the first observational trend is compatible with our simulations for late ages when we consider all planets inside an orbital period larger than 6~day (see Fig.~\ref{number_inside_Pdays_legend}). 
However, for early ages, we would expect the opposite. 
Fig.~ \ref{pop_exoplanet_eu_per_vs_met} shows also the age of the stars. 
It seems that the difference between metal poor and metal rich stars is less visible for earlier ages, however it is difficult to see a clear trend with age. 

The explanation of the second observational trend also appears to be compatible with our simulations for very late ages. 
While early in the history, close-in planets are located closer around the metal poor star (see bottom panel of Fig.~\ref{number_inside_Pdays_legend}), the difference decreases with time and at 8~Gyr the planets are actually located farther from the metal poor stars than around the metal rich stars. 
This reversal of trends is once again due to the fact that the planets around metal poor stars keep falling onto the star on much longer timescales than the planets around metal rich stars (see top panel of Fig.~\ref{number_inside_Pdays_legend}).
The bottom panel of Fig.~\ref{number_inside_Pdays_legend} also hints at a second order effect: we find that planets around the metal rich star are found slightly farther away than the planets around the solar metallicity star.
This is mainly due to the fact that the planets around the initially fast rotating metal rich stars migrate farther away than those around the initially fast rotating solar metallicity stars, while the fate of the planets around the initially moderately to slowly rotating stars is very similar for both metallicities (see Fig~\ref{model1days} and \ref{model3days}).
Interestingly enough, this second order trend is actually somewhat visible in the observational data seen in Fig.~\ref{pop_exoplanet_eu_per_vs_met} (black arrow).

To summarize, \textit{given our hypothesis about the initial planet population distribution and assuming the stars in the sample are old stars (older than 7~Gyr), our model allows us to reproduce the main observational trends of the distribution of hot Jupiters with the metallicity of their host star}.  
While this might be a happy coincidence, our results do underlie the fact that the mechanisms of tidal dissipation in stars probably play a major role in the shaping of the hot Jupiter population, and that these processes depend on metallicity and on rotation.   

\medskip

On the other hand, the theoretical works about planetary formation of \citet{Johnson2012} showed that there is a minimum metallicity required for planetary formation and that planets that form around metal poor stars should be found closer to the star after the protoplanetary disk dispersal.
Contrary to \citet{Johnson2012}, \citet{Mordasini12} showed that there is no clear correlation between metallicity and semi-major axis at the end of the planetary formation phase.
This statement of \citet{Mordasini12} corresponds to our hypothesis of a metallicity-independent initial semi-major axis distribution (hypothesis leading to Figs.~\ref{Pop_Age_infl_Z} and \ref{number_inside_Pdays_legend}).

Other works using models of planetary formation \citep{IdaLin2004,IdaLin2008} or models of population synthesis \citep{Mordasini2009,Mordasini12} also reproduced the fact that more planets are expected to be found around metal rich stars.
If we had considered an initial population of planets depending on the metallicity of the star, our simulation results would agree with the observational data for a wider range in ages. 
In other words, the reversal of the trend that we observe in the dependence of the number of close-in planets with metallicity between early and late ages (see Fig.~\ref{number_inside_Pdays_legend}) would occur much earlier.

Taking into account an initial population depending on stellar metallicity would allow us to conclude that the statistical trends of Fig.~\ref{pop_exoplanet_eu_per_vs_met} visible in the population of hot Jupiters could be due to the dependence of the stellar dissipation with metallicity. 

However, we need to go deeper in the comparison of our simulations with the observational trends and in the disentangling of the different effects (like initial planetary population, mass and age of star).
Namely, the next step would be to use a population of hot Jupiters coming from formation models \citep[e.g.][]{Mordasini12} to compute the evolution of this population and see how well this compares to the observations.
We also intend to look at the dependence of our results on the time of the protoplanetary disk dispersal as it could vary with initial stellar rotation \citep[e.g.][]{GB15}, metallicity as well as other quantities, such as the magnetic field. 
Another step would be to conduct a single spectroscopic analysis to have consistent estimates of the metallicities for the stars of different spectroscopic surveys \citep[as could be done with iSpec,][]{BlancoCuaresma2014} focusing $1~\Msun$ stars with hot Jupiters.

\section{Conclusion}
\label{conc}

We showed that the metallicity of a low-mass star has a strong effect on the stellar parameters which strongly influence the tidal dissipation in their convective region.
For instance, for a Sun-like star on the MS, the frequency-averaged dissipation can vary by two orders of magnitude between a metal poor star (Z = 0.0040) and a metal rich star (Z=0.0255).
We find that on the PMS the dissipation of a metal poor star is bigger than the dissipation of a metal rich star. 
However, on the MS, it is the opposite: the higher the metallicity, the higher the tidal dissipation.
While this is true for a Sun-like star, the dissipation in a less massive star (0.4 $M_\odot$) shows a weaker dependency on metallicity.

This metallicity dependence of the tidal dissipation has an effect on the orbital evolution of the planets around Sun-like stars.
Using the orbital evolution model of \citet{Bolmont16}, we show that changing the metallicity leads to different evolutions (e.g., planets migrate farther out from an initially fast rotating metal rich star, see Fig.~\ref{sma_diff_Z}).
We also studied the evolution of a statistical population of Jupiter-mass planets at different initial orbital periods around $1~\Msun$ stars of different initial rotation periods and for three different metallicities.

We reproduce qualitatively the observational trends of the population of hot Jupiters with the metallicity of their host stars.
Namely, we reproduce that for old stars (older than 6~Gyr), the higher the metallicity the higher the number of close-in planets and the higher the metallicity the closer the planets.
However, more steps are needed to try to quantitatively fit our results to the observations.
First, we need to keep improving our tidal dissipation models, for instance by including the dissipation in the radiative region of the star due to tidal gravity waves.
Second, we need to use in our simulations an initial population of planets more representative and realistic than the one we used \citep[coming from planetary formation models, e.g.][]{Mordasini12}.
Third, we need to make sure the estimations of the metallicities are done consistently in the observations (so that they are not just an accumulation of estimates from different surveys done with different methods, see discussion in \citealt{BlancoCuaresma2016}).

TESS \citep{Campante2016} and PLATO \citep{Plato} will bring the community invaluable informations to confront our statistical approach for the tidal evolution of close-in Jupiters. 
Our simulations outputs together with a better estimation of the dependence of the initial planet population with the stellar parameters (metallicity, rotation period, time of disk dispersal...) would allow us to make predictions on the population of hot Jupiters for different ages.
For instance, with our assumptions, we can see different trends for the population of hot Jupiters around initially fast rotating stars vs initially slow rotating stars, around metal poor stars vs metal rich stars as a function of the age of the star.
Another example, is that we can also infer from all our simulations that to have 1\% of hot Jupiters observed around solar mass MS stars \citep[as observational surveys tend to show, e.g.][]{Marcy2005,Cumming2008,Mayor2011,Wright2012}, the fraction of hot Jupiters after the protoplanetary disk dispersal should be about 1.8\%.
With the capacity of TESS and PLATO to estimate the stellar ages and masses with asteroseismology, we will have an unequaled insight into the orbital evolution of close-in planets.

\begin{acknowledgements}
The authors would like to thank the anonymous referee for his/her constructive comments.
E. B. acknowledges that this work is part of the F.R.S.-FNRS ExtraOrDynHa research project.
S.M. and E.B. acknowledge funding by the European Research Council through ERC grant SPIRE 647383. 
The authors acknowledge financial support from the Swiss National Science Foundation (FNS) and from the French Programme National National de Physique Stellaire PNPS of CNRS/INSU. 
This work results within the collaboration of the COST Action TD 1308.
This work has been carried out in part within the frame of the National Centre for Competence in Research PlanetS supported by the Swiss National Science Foundation.
\end{acknowledgements}

%\bibliographystyle{aa}
%\bibliography{Reference}

 \newcommand{\noop}[1]{}

\end{document}